\documentclass[acmsmall]{acmart}

\usepackage[utf8]{inputenc}

\usepackage{todonotes}
\usepackage{nth}

\usepackage{multirow}

\AtBeginDocument{%
  \providecommand\BibTeX{{%
    \normalfont B\kern-0.5em{\scshape i\kern-0.25em b}\kern-0.8em\TeX}}}

\setcopyright{none}
\copyrightyear{2022}
\acmYear{2022}
\acmPrice{}
\acmDOI{10.1145/3558052}

\acmJournal{CSUR}
\acmVolume{99}
\acmNumber{9}
\acmArticle{999}
\acmMonth{8}

\graphicspath{graphics/}
\DeclareGraphicsExtensions{.jpg}

\usepackage[normalem]{ulem}

\begin{document}

\title{A Survey of Security and Privacy Issues in V2X Communication Systems} 

{
\noindent
\footnotesize This is an extended version of an article to appear in ACM
Computing Surveys (CSUR) with the same title in August 2022 under
\url{https://dl.acm.org/doi/10.1145/3558052}.
}

\author{Takahito Yoshizawa}
\affiliation{%
  \institution{imec-COSIC KU Leuven}
  \streetaddress{Kasteelpark Arenberg 10 Bus 2452}
  \city{Leuven}
  \postcode{B-3001}
  \country{Belgium}
}
\email{takahito.yoshizawa@esat.kuleuven.be}
\orcid{0000-0001-5684-9597}
%\authornotemark[1]

\author{Dave Singelée}
\affiliation{%
 \institution{imec-COSIC KU Leuven}
 \streetaddress{Kasteelpark Arenberg 10 Bus 2452}
 \city{Leuven}
 \postcode{B-3001}
 \country{Belgium}
}
\email{dave.singelee@esat.kuleuven.be}

\author{Jan Tobias M\"uhlberg}
\affiliation{%
  \institution{imec-DistriNet, KU Leuven}
  \streetaddress{Celestijnenlaan 200a box 2402}
  \city{Leuven}
  \postcode{B-3001}
  \country{Belgium}
}
\email{jantobias.muehlberg@cs.kuleuven.be}

\author{Stéphane Delbruel}
\affiliation{%
  \institution{LaBRI, University of Bordeaux}
  \streetaddress{Cours de la Libération, 351}
  \city{Talence}
  \postcode{F-33405}
  \country{France}
  }
\email{stephane.delbruel@labri.fr}

\author{Amir Taherkordi}
\affiliation{%
	\institution{Informatics Department,  University of Oslo}
	\streetaddress{Gaustadalléen 23B}
	\city{Oslo}
	\postcode{0373}
	\country{Norway}
}
\email{amirhost@ifi.uio.no}

\author{Danny Hughes}
\affiliation{%
  \institution{imec-DistriNet, KU Leuven}
  \streetaddress{Celestijnenlaan 200a box 2402}
  \city{Leuven}
  \postcode{B-3001}
  \country{Belgium}
  }
\email{danny.hughes@cs.kuleuven.be}

\author{Bart Preneel}
\affiliation{%
  \institution{imec-COSIC KU Leuven}
  \streetaddress{Kasteelpark Arenberg 10 Bus 2452}
  \city{Leuven}
  \country{Belgium}
}
\email{bart.preneel@esat.kuleuven.be}

\renewcommand{\shortauthors}{Yoshizawa, et al.\ }

\begin{abstract}
Vehicle-to-Everything (V2X) communication is receiving growing attention from
industry and academia as multiple pilot projects explore its capabilities
and feasibility. With about 50\% of global road vehicle exports coming from
the European Union (EU), and within the context of EU legislation around security and
data protection, V2X initiatives must consider security and privacy aspects
across the system stack,  in addition to road safety. Contrary to this
principle, our survey of relevant standards, research outputs, and EU pilot
projects indicates otherwise; we identify multiple security and privacy
related shortcomings and inconsistencies across the standards. We conduct
a root cause analysis of the reasons and difficulties associated with these
gaps, and categorize the identified security and privacy issues relative to these root 
causes. As a result, our comprehensive analysis sheds lights on a number of areas 
that require improvements in the standards, which are not explicitly
identified in related work.
Our analysis fills gaps left by other related surveys, which
are focused on specific technical areas but not necessarily point out underlying
root issues in standard specifications. We bring forward recommendations to address these gaps 
for the overall improvement of security and safety in vehicular communication.
\end{abstract}


%%
%% Keywords. The author(s) should pick words that accurately describe
%% the work being presented. Separate the keywords with commas.

\vspace{3mm}
\maketitle

% !TEX root = v2x_survey_main.tex
% !TeX spellcheck = en-GB

% ========  Section 1  ==========
\section{Introduction} 
\label{sec:introduction}
\subsection{Overview}
\label{sec:intro_overview}
Vehicular communication has gained momentum in the past years. 
It is expected that progress of this communication technology will have a major impact on the automotive industry and in particular on how vehicles are driven in society.
Vehicle-to-Everything (V2X) communication is expected to bring numerous benefits. 
According to the assessment by National Highway Traffic Safety Administration (NHTSA) in the US~\cite{nhtsa2016}, the adoption of Vehicle-to-Vehicle (V2V) technology is expected to improve the overall traffic safety by preventing 439,000 to 615,000 accidents, saving 987 to 1366 lives, and eliminating 537,000 to 746,000 property damage incidents per year.
A report by the European Commission (EC)~\cite{asselin2016study} states that the overall benefits of deploying Cooperative Intelligent Transport System (C-ITS) include reduced travel times, increased efficiency, reduced accident rates, and savings in fuel consumption.

\begin{figure}[h]
  \centering\includegraphics[width=.9\linewidth]{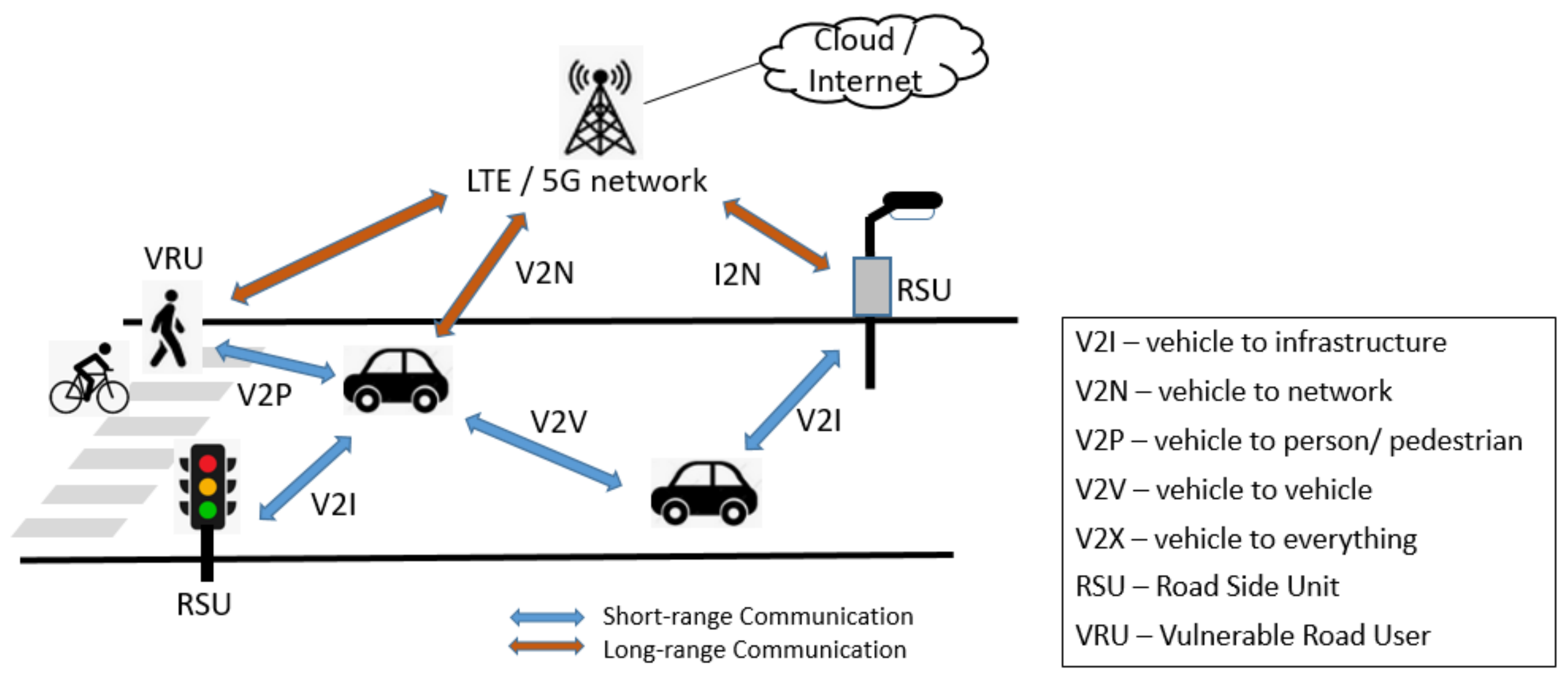}
  \vspace{-0.2cm}
  \caption{Overview of V2X Communication System}
  \label{fig:OverviewV2XCommSystem}
\end{figure}

The first standardized V2X technology is based on IEEE 802.11p (IEEE 802.11 Outside the Context of Basic Service Set (OCB) mode). 
In the US, the V2X system using 802.11 OCB mode is called Dedicated Short-Range Communication (DSRC) and its upper layer is called Wireless Access in Vehicular Environment (WAVE) as specified in the IEEE 1609 series and the SAE International (SAE) standard J2735~\cite{saeJ2735} (formerly the Society of Automotive Engineers). 
In Europe, ITS systems based on IEEE 802.11 OCB mode are called Intelligent Transport System G5 (ITS-G5), and its upper layer is referred to as C-ITS\@. 
Detailed descriptions of these standards and how they relate to each other are discussed in~\cite{yoshizawa2019survey}. 
An overview of the V2X communication system is shown in Figure~\ref{fig:OverviewV2XCommSystem}. 
V2X is a collective term which includes multiple communication modes:

\begin{itemize}
	\item Vehicle-to-Vehicle (V2V): direct communication between vehicles.
	\item Vehicle-to-Infrastructure (V2I): communication between a vehicle and the infrastructure such as traffic light.
	\item Vehicle-to-Person/Pedestrian (V2P): between a vehicle and
other road users such as pedestrians or cyclists.
	\item Vehicle-to-Network (V2N): between a vehicle and network entities via a mobile network base station.
	\item Infrastructure-to-Network (I2N): between an infrastructure and network entities via a mobile network.
\end{itemize}

Meanwhile, modern vehicles are equipped with increasing number of
Electronic Control Units (ECUs), embedded computers with the power to sense
and actuate within the vehicle. 
Today, standard automobiles have more than 80 ECUs; luxury ones have as many as 150 ECUs~\cite{syrmasgsurl,mckinseyurl}. Their software size amounts to 100 million lines of code (LoC), far exceeding that of the space shuttle (400,000~LoC), F35 fighter jet (23~millin~LoC) and even the Hadron Collider (50~million~LoC)~\cite{iotworldtoday2022}. 
The increased complexity of vehicular ICT---ECUs, connectivity, overall
software size---implies an overall increasing digital attack surface and
thereby increasing security and privacy risks.  
Contrary to this point, the Controller Area Network (CAN) bus in
vehicles does not support protection against cyberattacks; it does not
support authentication and message integrity
functionalities~\cite{mun2020ensuring}. In fact, the CAN design is based on
the assumption that it operates in a friendly environment where no security
threats exist~\cite{bella2019toucan}. Yet, real-life attack scenarios
around CAN communication and software defects in ECUs and their consequence
as potential road hazard raise immediate
concerns~\cite{koscher2010experimental,checkoway2011comprehensive,miller:remote_car_exploit,mun2020ensuring}.
Even if a relatively small number of vehicles on the road are directly affected by
an attack, the consequences for road safety and national infrastructures
can be devastating~\cite{vivek2019cyberphysical}. Therefore, vehicular communication
technologies need to be secure, robust, and resilient to be truly
beneficial. 
In this sense, including security and safety requirements from the initial stage of the system design is essential~\cite{mun2020ensuring}. 
Ensuring secure communication requires at least
mechanisms for authentication and integrity protection, which allows
communicating parties to verify each other's identity, and the authenticity
of messages: a vehicle is indeed a vehicle and a traffic light is indeed a
traffic light, and communication between these entities cannot be spoofed
or otherwise manipulated.
In addition, protecting the \textit{privacy} of vehicle owners is equally important. If
no privacy protections are implemented on top of authenticated
communications, vehicles can be tracked remotely, and information about
drivers and their personal behaviour can be inferred by authorities,
infrastructure operators and adversaries. In the V2X context,
privacy-preserving technologies rely on the use of pseudonymous identities.
These schemes come with their own security and safety
challenges~\cite{rouf2010security,lefevre2013privacy}, and a number of
proposals to mitigate these challenges have been
published~\cite{muhammad2018survey,brecht2018security,lu2020v2xreview}, but
not necessarily adopted by standardizing bodies.

\begin{figure}[t]
	\centering
	\vspace{-0.3cm}
	\includegraphics[width=\linewidth]{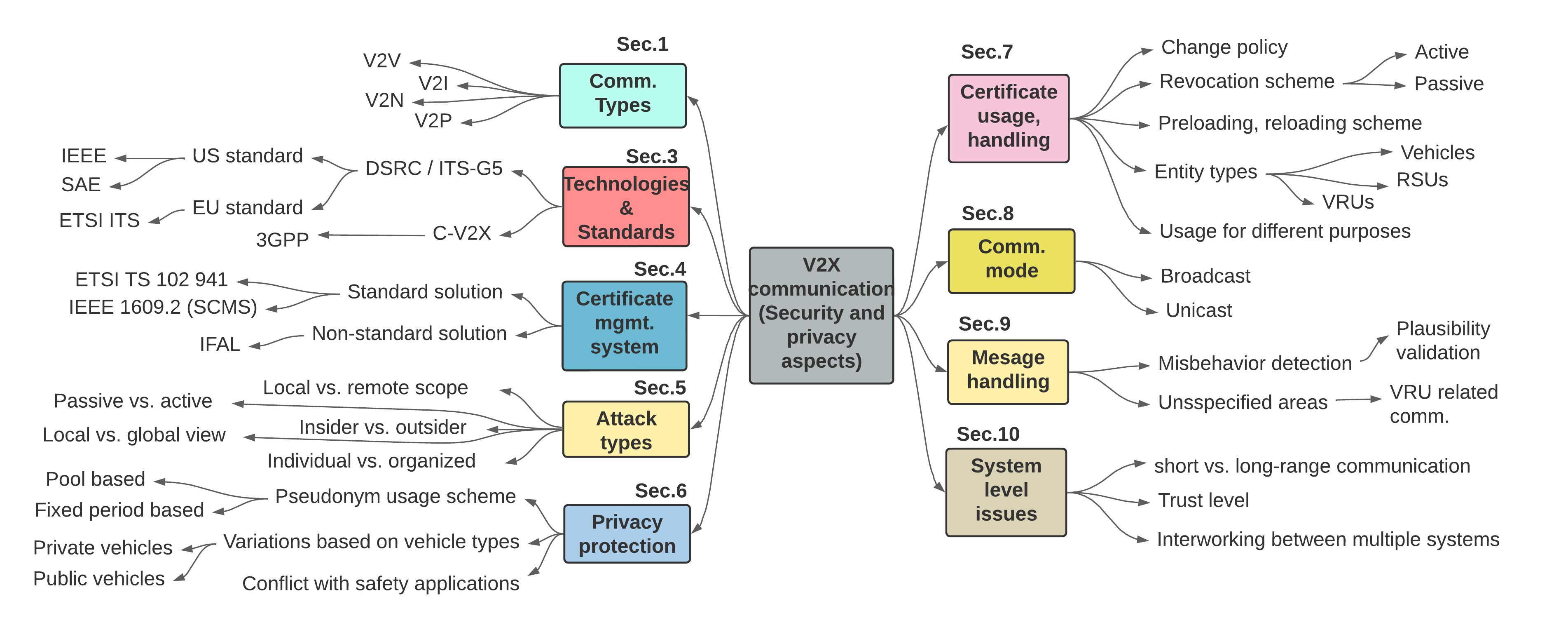}
	\vspace{-0.8cm}
	\caption{Overview of Security and Privacy Aspects in V2X Systems.}
	\vspace{-0.2cm}
	\label{fig:OverviewSecPriElements}
\end{figure}

This article surveys recent efforts in research and standardization
regarding security and privacy aspects of V2X
communication. We aim to provide a comprehensive overview of these aspects
from the perspective of the ETSI ITS specifications, and link this to the
objectives and results of recent research, development, and integration
actions, specifically in the European Union but with links to other markets
and political bodies. As a key contribution, our survey highlights a range
of shortcomings and imprecisions in ongoing standardization efforts, and
points towards potential solutions and open research questions regarding
these security and privacy issues. As such, our article targets policy
makers, researchers, and standardization engineers. Specifically for these
actors, we provide the background to assess the novelty and relevance of
research, and a medium to identify domains of research with a potentially
high impact on upcoming standardization or regulatory efforts.
Figure~\ref{fig:OverviewSecPriElements} gives an overview of these domains
and links them to the structure of this article.

\subsection{Methodology and Contributions}
\label{sec:intro_methodology}

In this article, and in reference to related work summarized in
Sec.~\ref{sec:related_work}, we focus on security and privacy aspects of
European Union (EU) initiatives across a range of European pilot
activities, such as past or present V2X-related EU projects, and analyse
ETSI ITS specifications from security and privacy perspectives. 
Our focus on EU standard and projects is due to several reasons: (1) Europe
is the largest exporter of vehicles equipped with communication
technologies as many major vehicle manufacturers are based in Europe, (2)
covering relevant projects in other regions such as the US will diminish
the focused analyses, and (3) including analysis of other regions will
likely result in excessively long paper while adding relatively marginal
additional benefit.

As a result, we identify a number of gaps in the ETSI ITS standards which
stem from incoherent and inconsistent specifications. Based on our root cause analysis 
of the reasons and difficulties associated with these gaps, we provide
recommendations based on our findings. We also identify previously
unreported security and privacy issues resulting from recent trends driven
by C-V2X such as the convergence of vehicular communication and mobile
systems; one notable example being the absence of security and privacy
considerations in the integration of smartphone in the ITS system in V2P
scenarios. For many of these issues, existing literature offers
either no solution or inadequate ones. 
When applicable, we propose ideas for potential solutions as an agenda for
future analysis and investigation. We believe our findings are fundamental
and essential to ensure that V2X communication is secure and protect user
privacy.

We focus primarily on the EU initiatives in this paper. However, subtle but
important contrast emerges as we put differences of the solutions between
the EU and the US in perspective because the security and privacy solutions
in two systems are similar in high level but different in details.  For
this reason, we discuss the similarities and differences between the two
standards as the foundation of our discussion.  When terminologies are
different between the systems, we use the EU terminology unless it is
necessary to explicitly describe the US system. 

\subsection{Document Organization} \label{sec:intro_organizatin}
This paper is organized as follows. 
We first set a baseline of the discussion by summarizing the related survey papers and EU projects in Sec.~\ref{sec:related_work}, describing the overview of V2X technologies and their standards in Sec.~\ref{sec:v2xTechnologies}. 
Then, we discuss their security management systems in Sec.~\ref{sec:systemOverview}. 
In Sec.~\ref{sec:threatModel}, we discuss threat model that we use as a base of our analysis. 
Starting from Sec.~\ref{sec:issuePrivacyProtection} and up to Sec.~\ref{sec:issueSystemLevel}, we examine details of specific security aspects and gaps. 
In Sec.~\ref{sec:rootCauseAnalysis}, we present our root cause analysis of key gaps that need further study and research, and put forward recommendations.
We conclude our study in Sec.~\ref{sec:conclusion}. 
As a guide for the rest of this paper, Figure~\ref{fig:OverviewSecPriElements} provides an illustration of the various security and privacy protection-related aspects covered in this paper.

% !TEX root = v2x_survey_main_arxiv.tex
% !TeX spellcheck = en-GB

% ========  Section 2 (Option_2)  ==========

\section{Related Work: Extended Overview}
\label{sec:related_work}
\subsection{Survey Papers}
\label{sec:related_work_papers}

\begin{table}[!htbp]
	\footnotesize
	\vspace{-0.2cm}
	\caption{V2X RELATED SURVEY PAPERS}
	\vspace{-0.3cm}
	\label{tab:Survey_papers}
	\begin{tabular}{lcccll}
		\hline
		\noalign{\hrule height 1pt}
		Author & Year & Ref. & Group & Focused area & High level description  \\ \hline
		\noalign{\hrule height 1pt}
		Hasan et al. 
		& 2020
		& \cite{hasan2020securing}
		& \raisebox{.5pt}{\textcircled{\raisebox{-.6pt} {1}}}
		& \begin{tabular}[c]{@{}l@{}} misbehaviour \\detection \end{tabular}  
		& \begin{tabular}[c]{@{}l@{}} Analysis of attack types and misbehaviour detection \\
			mechanisms. \end{tabular} 
		\\ \hline
		\begin{tabular}[c]{@{}l@{}} van der Heijden \\et al. \end{tabular} 
		& 2018
		& \cite{vanderheijden2018survey} 
		& \raisebox{.5pt}{\textcircled{\raisebox{-.6pt} {1}}}
		& \begin{tabular}[c]{@{}l@{}} misbehaviour \\detection \end{tabular} 
		& \begin{tabular}[c]{@{}l@{}} In-depth analysis of misbehaviour detection \\
			mechanisms and classifications. \end{tabular} 
		\\ \hline
		Bi{\ss}meyer 
		& 2014
		& \cite{bissmeyer2014thesis} 
		& \raisebox{.5pt}{\textcircled{\raisebox{-.6pt} {1}}}
		& \begin{tabular}[c]{@{}l@{}} misbehaviour \\detection \end{tabular} 
		& \begin{tabular}[c]{@{}l@{}} Misbehaviour detection and attacker identification \\
			in both local short-term and central long-term scope. \end{tabular} 
		\\ \hline
		Badea and Stanciu  
		& 2018
		& \cite{badea2018survey} 
		& \raisebox{.5pt}{\textcircled{\raisebox{-.6pt} {1}}}
		& car sharing
		& \begin{tabular}[c]{@{}l@{}} Limited coverage on security aspects, focused on \\
			specific use cases and scenarios, e.g. car sharing. \end{tabular} 
		\\ \hline
		\begin{tabular}[c]{@{}l@{}} Ometov and \\Bezzateev \end{tabular} 
		& 2017
		& \cite{ometov2017multi} 
		& \raisebox{.5pt}{\textcircled{\raisebox{-.6pt} {1}}}
		& MFA 
		& \begin{tabular}[c]{@{}l@{}} Despite its title, it is not a survey; it is a proposal to \\
			apply multi-factor authentication (MFA) in V2X. \end{tabular} 
		\\ \hline
		Wang et al. 
		& 2020
		& \cite{wang2020certificate}
		& \raisebox{.5pt}{\textcircled{\raisebox{-.6pt} {1}}}
		& \begin{tabular}[c]{@{}l@{}} certificate \\revocation \end{tabular} 
		& \begin{tabular}[c]{@{}l@{}} Analysis of various revocation schemes and classify \\
			them based on the location where the revocation \\
			information has been placed. \end{tabular}  
		\\ \hline
		Huang et al.
		& 2020
		& \cite{huang2020recent} 
		& \raisebox{.5pt}{\textcircled{\raisebox{-.6pt} {2}}}
		& \begin{tabular}[c]{@{}l@{}} security and \\privacy solutions \end{tabular}
		& \begin{tabular}[c]{@{}l@{}} Categorization of security and privacy solutions \\
			from two  perspectives: 1) cryptography-based \\
			schemes, and 2) trust-based schemes. Discussion on \\
			both LTE and 5G-based C-V2X security.\end{tabular}
		\\ \hline
		Alnasser et al.
		& 2019
		& \cite{alnasser2019cyber}   
		& \raisebox{.5pt}{\textcircled{\raisebox{-.6pt} {2}}}
		& \begin{tabular}[c]{@{}l@{}} security threats \\and solutions \end{tabular}
		& \begin{tabular}[c]{@{}l@{}} Analysis of security threats and solutions for both \\
			DSRC and LTE-based C-V2X. Threat analysis includes \\
			availability, integrity, confidentiality authenticity, \\
			and non-repudiation. \end{tabular} 
		\\ \hline
		Ghosal and Conti 
		& 2020
		& \cite{ghosal2020security} 
		& \raisebox{.5pt}{\textcircled{\raisebox{-.6pt} {2}}}
		& \begin{tabular}[c]{@{}l@{}} attack types\\ and solutions \end{tabular}
		& \begin{tabular}[c]{@{}l@{}} Analysis on security challenges and requirements \\
			by examining various attack types, their impacts, \\
			followed by solution analysis. \end{tabular}
		\\ \hline
		Cao et al. 
		& 2019
		& \cite{cao2019survey}  
		& \raisebox{.5pt}{\textcircled{\raisebox{-.6pt} {3}}}
		& \begin{tabular}[c]{@{}l@{}} 5G system \end{tabular} 
		& \begin{tabular}[c]{@{}l@{}} Covers overall 5G system security and privacy \\
			aspects of vertical applications (e.g. IoT, D2D, and \\
			5G-specific topics such as network slice) rather than \\
			specifically on 5G-based C-V2X. \end{tabular} 
		\\ \hline
		\begin{tabular}[c]{@{}l@{}} Muhammad \\and Safdar \end{tabular} 
		& 2018
		& \cite{muhammad2018survey}  
		& \raisebox{.5pt}{\textcircled{\raisebox{-.6pt} {3}}}
		& \begin{tabular}[c]{@{}l@{}} LTE C-V2X \end{tabular} 
		& \begin{tabular}[c]{@{}l@{}} Focus is specifically on the authentication mechanism \\
			of LTE systems in C-V2X context. They describe \\
			attack types, their relevance and countermeasures \\
			within the C-V2X context. \end{tabular} 
		\\ \hline
		Lu et al. 
		& 2019
		& \cite{lu2020v2xreview} 
		& \raisebox{.5pt}{\textcircled{\raisebox{-.6pt} {3}}}
		& \begin{tabular}[c]{@{}l@{}} LTE, \\5G C-V2X \end{tabular} 
		& \begin{tabular}[c]{@{}l@{}} Challenges in trust, security, and privacy-related \\
			issues in C-V2X, and strategies to resolve each of \\
			them for LTE and 5G-based C-V2X. \end{tabular} 
		\\ \hline
		Lai et al. 
		& 2020
		& \cite{lai2020security}
		& \raisebox{.5pt}{\textcircled{\raisebox{-.6pt} {3}}}
		& \begin{tabular}[c]{@{}l@{}} platoon \\operation \end{tabular} 
		& \begin{tabular}[c]{@{}l@{}} Proposal of security solutions specifically for platoon \\
			operation, including privacy-preserving group set up, \\
			distributed group key management, and cooperative \\
			message authentication. \end{tabular} 
		\\ \hline
		Marojevic 
		& 2018
		& \cite{marojevic2018c}
		& \raisebox{.5pt}{\textcircled{\raisebox{-.6pt} {3}}}
		& \begin{tabular}[c]{@{}l@{}} LTE C-V2X \end{tabular} 
		& \begin{tabular}[c]{@{}l@{}} Focus is on threat scenarios of LTE-based C-V2X, \\
			lists out associated security requirements, and sets \\
			out research directions to satisfy these requirements \\
			along with the needs of further standardization. \end{tabular} 
		\\ \hline
		\noalign{\hrule height 1pt}
	\end{tabular}
\end{table}

We first review related work of V2X communication. %\footnote{This is an extended version of Sec.2 in the main paper.}
In the past years, a number of survey papers on security and privacy
aspects of V2X have been published. 
Table~\ref{tab:Survey_papers} captures the overview of these surveys. 
The purpose of this section is to highlight the differences of our work compared to them.  
These survey papers can be categorized into three
groups: \raisebox{.5pt}{\textcircled{\raisebox{-.6pt} {1}}} discussion and solutions for generic vehicular network,
\raisebox{.5pt}{\textcircled{\raisebox{-.6pt} {2}}}
surveys that cover both DSRC and C-V2X based solutions, and \raisebox{.5pt}{\textcircled{\raisebox{-.6pt} {3}}} surveys that focus specifically on C-V2X solutions. 

In the first group, Hasan et al.~\cite{hasan2020securing}, van der Heijden et al.~\cite{vanderheijden2018survey}, and Bi{\ss}meyer~\cite{bissmeyer2014thesis} exclusively focus on misbehaviour detection. 
In~\cite{hasan2020securing}, a large part of this survey is dedicated to the discussion on attack and misbehaviour detection mechanisms by classifying and comparing numerous proposals on this subject. 
Survey in~\cite{vanderheijden2018survey} provides in-depth analysis of detection mechanisms, and classify them into two-dimensions. The first dimension is \emph{node-centric vs. data-centric}; the second dimension is \emph{autonomous vs. collaborative}. Based on this classification, misbehaviour detection mechanisms can be categorized into one of the four types. The work in ~\cite{bissmeyer2014thesis} focuses in two main areas: 1) misbehaviour detection, and 2) attacker identification in both local short-term and central long-term scope.

The survey by Badea and Stanciu~\cite{badea2018survey} has a limited focus on security and privacy aspects. The only security related topic is on specific use cases and scenarios such as car sharing. 
Similarly, Ometov and Bezzateev~\cite{ometov2017multi} is not a survey; it is a proposal to apply multi-factor authentication (MFA) in V2X communication. This solution uses reversed Lagrange polynomial from Shamir's secret sharing schema as the building block. 
Wang et al.~\cite{wang2020certificate} focus specifically on the
certificate revocation schemes in V2X communication. They analyse and classify them based on location where the
revocation information has been placed. The entire revocation process is
then divided into three stages: 1) resolution, 2) distribution, and 3) the use of revocation information.

In the second group, Huang et al.~\cite{huang2020recent} treat both security and privacy topics although it is a small part of their paper as the content provides overview of 3GPP specifications on both LTE and 5G-based C-V2X rather than a survey. For security aspect, they discuss basic security services and attack types in vehicular networks, then categorize security solutions from two perspectives: 1) cryptography-based schemes, and 2) trust-based schemes. For privacy solutions, they categorize them into two types: 1) identity privacy preservation, and 2) location privacy preservation.
Alnasser et al.~\cite{alnasser2019cyber} analyse security threats and solutions for both DSRC and LTE-based C-V2X. Their threat analysis includes availability, integrity, confidentiality, authenticity, and non-repudiation aspects. Then, they categorize security solutions including cryptography-based, behaviour/trust-based, and identity-based solutions. 
Ghosal and Conti~\cite{ghosal2020security} provide overview and background of both DSRC/WAVE and 3GPP-based LTE and 5G C-V2X systems, then discuss security challenges by analysing various attack types and how they impact V2X communication. Later, they examine techniques and solutions in specific areas, including symmetric key cryptography, privacy preservation, message authentication.

In the third group, Cao et al.~\cite{cao2019survey} cover overall 5G system security and related topics rather than focusing on 5G-based C-V2X. They describe security and privacy aspects of other vertical applications such as IoT, device-to-device (D2D), and 5G-specific topics such as network slice. Only one section (Sec.VI) is dedicated to security in both LTE and 5G-based C-V2X solutions. In it, security requirements, solutions, and open issues are discussed. 
Muhammad and Safdar~\cite{muhammad2018survey} focus specifically on the authentication mechanism in the context of LTE C-V2X. They enumerate attack types and describe their relevance and corresponding countermeasures. They discuss multiple proposed authentication solutions and analyse how they can meet the needs of C-V2X communication.
Lu et al.~\cite{lu2020v2xreview} focus on security and privacy aspects of LTE and 5G-based C-V2X. Their survey is on challenges in trust, security, and privacy-related issues in C-V2X, followed by discussions on strategies to resolve them. 
The discussion in Lai et al.~\cite{lai2020security} rather narrowly focuses on a specific C-V2X scenario (platooning) rather than more general V2X communication. They propose security solutions, including privacy-preserving platoon group set up, distributed group key management, and cooperative message authentication.
Marojevic~\cite{marojevic2018c} focuses on LTE-based C-V2X, discusses its threat scenarios, lists out associated security requirements, and proposes research directions to satisfy these requirements, along with the needs of further standardization to ensure security mechanisms are in place.
In summary, the examined surveys can be characterized as follows: 
\vspace{-0.1cm}
\begin{itemize}
	\item Analysis of relevant threat scenarios and attack types and their impacts to V2X communication,
	\item Discussion of issues, challenges, and requirements on specific security areas,
	\item Discussion on approaches and strategies to address stated issues and requirements,
	\item Description and comparison of existing proposed solutions on specific security areas,
	\item A new proposal on specific security or privacy area.
\end{itemize}

In contrast to these surveys, as articulated in Sec.1.2, our unique contribution in this paper is to point out fundamental issues and gaps in the ETSI ITS standards and unreported security and privacy issues that have not been discussed in the existing survey papers.

\vspace{-0.1cm}
\subsection{European Union Research Actions}
\label{sec:related_work_projects}

\begin{table}[h]
	\footnotesize
	\vspace{-0.4cm}
	\caption{V2X RELATED EU PROJECTS}
	\vspace{-0.3cm}
	\label{tab:EUprojects}
	\begin{tabular}{llll}
		\hline
		\noalign{\hrule height 1pt}
		Project Name & Full Project Name   & Project Period   & Ref.      \\ \hline
		\noalign{\hrule height 1pt}
		PREVENT & Preventive and Active Safety Application & 2004-02$\sim$2008-01 & \cite{preventUrl} \\ \hline
		SeVeCom & SEcure VEhicle COMmunication & 2006-01$\sim$2008-12 & \cite{sevecomUrl}\\ \hline
		PRECIOSA & \begin{tabular}[c]{@{}l@{}} Privacy Enabled Capability In CO-operative systems and \\
			Safety Applications \end{tabular} & 2008-03$\sim$2010-08 & \cite{preciosaUrl} \\ \hline
		INTERSAFE-2 & Cooperative Intersection Safety & 2008-06$\sim$2011-05 & \cite{interSafe2Url} \\ \hline
		EVITA & E-safety Vehicle Intrusion proTected Applications & 2008-07$\sim$2011-12 & \cite{evitaUrl}\\ \hline
		OVERSEE & Open VEhiculaR SEcurE platform & 2010-01$\sim$2012-12 & \cite{overseeUrl}\\ \hline
		EcoGem & \begin{tabular}[c]{@{}l@{}} Cooperative Advanced Driver Assistance System for
			Green Cars \end{tabular} & 2010-09$\sim$2013-02 & \cite{ecogemUrl} \\ \hline
		PRESERVE & Preparing Secure Vehicle-to-X Communication Systems & 2011-01$\sim$2015-06 & \cite{preserveUrl} \\ \hline
		CONVERGE & Communication Network Vehicle Road Global Extension & 2011-08$\sim$2015-10 & \cite{convergeUrl} \\ \hline
		ICSI & Intelligent Cooperative Sensing for Improved Traffic Efficiency & 2012-11$\sim$2015-04 & \cite{icsiUrl} \\ \hline
		Autonet 2030 & \begin{tabular}[c]{@{}l@{}} Co-operative Systems in Support of Networked Automated \\
			Driving by 2030 \end{tabular}  & 2013-11$\sim$2016-10 & \cite{autonet2030Url} \\ \hline
		HIGHTS & High precision positioning for cooperative ITS applications & 2015-05$\sim$2018-04 & \cite{hightsmUrl} \\ \hline
		SCOOP@F Part2 & SCOOP & 2016-01$\sim$2018-12 & \cite{scoopUrl} \\ \hline
		ADASANDME & \begin{tabular}[c]{@{}l@{}} Adaptive ADAS to support incapacitated drivers Mitigate \\
			Effectively risks through tailor made HMI under automation \end{tabular} & 2016-09$\sim$2020-02 & \cite{adasAndMeUrl} \\ \hline
		KoMoD & Kooperative Mobilit{\"a}t im digitalen Testfeld D{\"u}sseldorf & 2017-06$\sim$2019-06  & \cite{koMoDUrl} \\ \hline
		5GCAR        & 5G Communication Automotive Research and innovation   & 2017-06$\sim$2019-07 & \cite{5gcarurl}  \\ \hline
		CONCORDA     & Connected Corridor for Driving Automation   & 2017-10$\sim$2020-06 & \cite{Concordaurl}  \\ \hline
		Synergy CAV & Synergy CAV- Connectivity and intelligent mobility & 2017-11$\sim$2020-04 & \cite{synergyUrl} \\ \hline
		PRoPART & Precise and Robust Positioning for Automated Road Transports & 2017-12$\sim$2019-11 & \cite{propartUrl} \\ \hline
		5G CARMEN    & \begin{tabular}[c]{@{}l@{}}5G for Connected and Automated Road Mobility in the \\
			European UnioN\end{tabular}   & 2018-11$\sim$2021-10 & \cite{5gcarmenurl}  \\ \hline
		5G CroCo     & 5G Cross Border Control   & 2018-11$\sim$2021-10 & \cite{5gcrocourl} \\ \hline
		5G-Mobix     & \begin{tabular}[c]{@{}l@{}}5G for Cooperative \& Connected Automated MOBility on \\
			X-border corridors\end{tabular} & 2019-06$\sim$2022-06 & \cite{5gmobixurl} \\ \hline
		C-Roads ITALY 3 & C-Roads ITALY 3 & 2020-07$\sim$2023-12 & \cite{CRoadsItalyUrl} \\ \hline
		\noalign{\hrule height 1pt}
	\end{tabular}
	\vspace{-0.1cm}
\end{table}

In Europe since 2008, there have been numerous EU projects that either
focus on or include V2X communications within their
scope~\cite{preventUrl,sevecomUrl,preciosaUrl,interSafe2Url,evitaUrl,overseeUrl,ecogemUrl,preserveUrl,convergeUrl,icsiUrl,autonet2030Url,hightsmUrl,adasAndMeUrl,koMoDUrl,5gcarurl,Concordaurl,synergyUrl,propartUrl,5gcarmenurl,5gcrocourl,5gmobixurl,CRoadsItalyUrl,scoopUrl},
which are listed in Table~\ref{tab:EUprojects} in chronological order. Most
of these projects are already completed at the time of writing and some of
them no longer have active websites. In such cases, we refer to the respective EU project websites~\cite{trimisUrl}, project deliverables, and research papers specific to these projects. Given that some of the most representative security related ETSI ITS specifications, such as ETSI TS 102 940~\cite{etsi2021102940} and 102 941~\cite{etsi2021102941}, first appeared in 2012, it is likely that some of the EU projects before that period had paved the way toward the creation of these security and privacy related specifications. Notable such examples are SeVeCom~\cite{sevecomUrl}, PRECIOSA~\cite{preciosaUrl}, EVITA~\cite{evitaUrl}, OVERSEE~\cite{overseeUrl}, and PRESERVE projects~\cite{preserveUrl}. 

SeVeCom project~\cite{sevecomUrl} addressed the privacy issue in V2V and V2I communication by defining a pseudonym change management mechanism to avoid the use of fixed MAC address which leads to positive identification of vehicles~\cite{kung2012ict}. PRECIOSA project~\cite{preciosaUrl} focused on the privacy protection of personal data in ITS applications. It adopted and clarified the privacy-by-design approach to address this issue~\cite{kung2012ict}. Objectives of EVITA project~\cite{evitaUrl} was to design, verify, and prototype security building blocks for automotive on-board networks to ensure sensitive data are protected against tampering. This project successfully designed a secure on-board architecture using Hardware Security Module (HSM) as the root-of-trust and secure on-board communication protocols~\cite{henniger2012evitaSummary}. OVERSEE project~\cite{overseeUrl} developed an open and secure vehicular platform that provides protected run-time environment that enables secure execution of multiple applications. This design protects networks both internal and external to the vehicle, using standardized interfaces, against active and passive attacks~\cite{oversee2013finalReport}. Finally, PRESERVE project~\cite{preserveUrl} specifically focused on security and privacy aspects of V2V and V2I communications as the primary scope of their work. It is likely that a large part of the security and privacy-related ETSI ITS specifications are the culmination of the results from these projects. Other projects not specifically mentioned up to this period ($\sim$2015)~\cite{preventUrl,interSafe2Url,ecogemUrl,convergeUrl,icsiUrl,autonet2030Url,hightsmUrl,adasAndMeUrl,koMoDUrl,5gcarurl,Concordaurl,synergyUrl,propartUrl,5gcarmenurl,5gcrocourl,5gmobixurl,CRoadsItalyUrl,scoopUrl} focus on other aspects of vehicular technologies and include V2X communication as a part of the overall project scope.

More recent EU projects primarily focus on demonstrations of technology readiness in V2X services in controlled or real-world environments rather than exclusively focusing on specific technical domains such as security or privacy aspects. It is especially notable in 5G Infrastructure Public Private Partnership (5G-PPP) related V2X projects such as in~\cite{5gcarurl,5gcarmenurl,5gcrocourl,5gmobixurl}. These projects are based on specific use cases and scenarios as described in the respective documents~\cite{naudts2018concordam19,fernandez20195gcard21,visintainer20195gcarmend21,perraud20195gcrocod21,martin20195gmobixd21}. Although these EU projects are beneficial to demonstrate the technology itself, none of them include security considerations within their scope. Studying their use case description documents reveals that almost all of them lack even a clear definition of underlying assumptions in the communication mode used (e.g.\ broadcast, unicast) or message direction (e.g.\ uplink, downlink). The underlying implicit assumption is that security solutions and mechanisms are in place, and all messages exchanged are verified, correct, and trustworthy. This assumption does not, however, reflects the reality. We have conducted a security and privacy gap analysis based on the aforementioned 5G-PPP EU projects, relevant ETSI ITS standard specifications, and research papers. As a result, we have identified significant security gaps and conflicts between the standard and what V2X communication aims to achieve. Our findings capture the main security and privacy gaps and concerns in V2X communication systems; they are confirmed by a number of other research papers~\cite{ghosal2020security,huang2020recent,alnasser2019cyber,chator2018don} that focus on the security aspects of V2X communication. Most notably is the work of Chator and Green~\cite{chator2018don}. In our analysis, we include the aspects specifically raised in~\cite{chator2018don} where relevant.

% !TEX root = ./v2x_survey_main.tex
% !TeX spellcheck = en-GB

% ========  Section 3  ==========

\section{V2X Technologies}
\label{sec:v2xTechnologies}
In this section, we discuss the two competing V2X standards, the first one being specified by the IEEE and the second by the 3GPP\@. 
We introduce these two main competing standards and highlight their differences as we set the context of the standardization in order to explore their respective security aspects and how they differ further in our survey. 
The discussion on their shared features being out of scope of this survey, we recommend for readers desirous to explore this path a previous study by Yoshizawa et al.~\cite{yoshizawa2019survey} focusing on the similarities of these two standards and possible hints on their cohabitation.

\subsection{IEEE standard}
\label{sec:v2xTechnologies_ss1}
In 2010, IEEE approved the amendment IEEE~802.11p designed to standardize vehicular communication system. The following publication of the IEEE~802.11 standard in 2016~\cite{ieee2016ieee80211} incorporated the amendment IEEE~802.11p. 
This amendment also specifies a new operation mode dubbed \textit{Outside Context of a Basic Service Set} (OCB) mode for each 802.11p compliant device to be set to. 
OCB mode does not need authentication nor association and the only parameter to set is the central channel frequency and the channel bandwidth to communicate. 
Overall, this amendment concerns the PHY and MAC layers for WLAN-based V2X communications. 
To build it up towards the applicative layer, the IEEE 1609 standard known as \textit{wireless access in vehicular environments} (WAVE) was developed by IEEE, while in Europe the ETSI committee on \textit{Intelligent Transportation Systems} (ETSI ITS) worked on top of IEEE 802.11p towards standardising applications and a security framework.
There are two initiatives benefiting from this work: SAE~\cite{saeJ2735}
is known as \textit{Dedicated Short-Range Communications} (DSRC), and ETSI ITS-G5.
Both of them define the upper layer protocols that operate on top of the 802.11 OCB mode. 
The intended application is short-range communication sufficient for direct communication involving both V2V between vehicles and V2I between vehicles and \textit{Road Side Units} (RSUs).

\subsection{3GPP standard}
\label{sec:v2xTechnologies_ss2}

Since 2014, the 3rd Generation Partnership Project (3GPP) has worked on standardizing vehicular communication based on 
previously standardised 4G LTE, and later on included 5G mobile cellular connectivity. 
This standardisation effort begun with the Proximity Services (ProSe) functionality published in Release-12 which was originally designed for public safety communication. Later in Release-13, support of direct communication between vehicles (D2D) was added.
To expand ProSe capabilities towards D2D communications in a cellular environment, a new interface called PC5 was defined in 3GPP TS 23.285~\cite{3gpp2019ts23285} for the LTE system. 
The equivalent functionality for the 5G system is in 3GPP TS 23.287~\cite{3gpp2020ts23287}.
The PC5 interface, also denoted \textit{sidelink}, facilitates a new communication path in addition to the existing \textit{Uu} interface between the User Equipment (UE) and the base station (the terminologies in standard specifications for base station in LTE and 5G are \textit{eNodeB} and \textit{gNodeB}, respectively). 
This combination of short-range sidelink (PC5) and long-range \textit{(Uu)} communications under the same system is considered complementary and enables a wide range of new types of use cases or services.
This technological approach relying on 4G LTE or 5G V2X communications is combined under the 3GPP standard for Cellular V2X (C-V2X).

\subsection{Differences between ITS-G5/DSRC and C-V2X}
\label{sec:v2xTechnologies_ss3}
In this section, we highlight conceptual differences between the two competing key V2X standards.
For the sake of clarity, the one carried by IEEE comprising IEEE 802.11~OCB mode along with DSRC/ETSI ITS-G5 will be designated using its ETSI denomination, i.e. ITS-G5, while the second defined by 3GPP will be specified under the term C-V2X.

\subsubsection{Two different target scenarios.}
\label{sec:v2xTechnologies_sss3_1}
The vehicular communication originally envisioned by the IEEE for the ITS-G5 was V2V where vehicles directly communicate with one another. 
Later on, communication that involves infrastructure, such as RSUs, was added. 
This type of communication is called V2I and in that scenario both V2V and V2I are designed from the point of view of the vehicle and its communications capabilities.

When the C-V2X concept emerged, it envisioned a scenario introducing two distinct new communication paradigms. 
One is the involvement of the cellular mobile network (V2N); the other is the involvement of pedestrians or cyclists through the use of smartphones (V2P). 
The addition of V2N and V2P paradigms is a natural extension to vehicle communication in the context of cellular mobile networks as both short-range and long-range communications in the C-V2X are defined by the same standard body (3GPP). 

Collectively, both ITS-G5-based systems and C-V2X-based systems use the term \textit{V2X}\@. 
However, the target scenarios are different. 
In the first one, the focus is on short-range communications involving vehicles and RSUs. 
Within the context of ITS-G5-based systems, the \textit{Intelligent
Transport System - Stations} (ITS-S) consists of only two types:
\textit{On-Board Unit} (OBU) and RSU\@. Thus, only dedicated devices for ITS are envisioned to be part of the vehicular communication system. 
In this sense, the term \textit{V2X} refers to V2V and V2I from the perspective of ETSI ITS-G5-based systems.

In the C-V2X-based system, inclusion of the mobile network (V2N) and pedestrians and cyclists (V2P) using a smartphone as a new type of ITS-S introduces new dimensions in the vehicular communication. 
Especially, introducing consumer devices within the family of ITS-S device types extend the range of communication options and use case scenarios. 
In this sense, the term \textit{V2X} refers to V2V, V2I, V2N, and V2P from the perspective of C-V2X-based systems.

\subsubsection{Specific Issues with ITS-G5}
\label{sec:v2xTechnologies_sss3_2}
Several concerns have been reported regarding the use of ITS-G5/DSRC within the context of vehicular communication. 
For example, Klingler et al.\ \cite{klingler2015ieee} show that the use of unicast in IEEE 802.11p OCB mode leads to a \textit{Head-of-Line blocking}\@ as each unicast frame requires an acknowledgment from the receiving ITS-S. Absence of acknowledgment in unicast causes subsequent frames to be blocked from transmission, not only to a specific unicast communication but also to other outbound traffic from the ITS-S in question. 
To create this \textit{Head-of-Line blocking} condition, only a minimally sophisticated attack is necessary as all it requires is to prevent the reception of an acknowledgment frame. 
Moreover, this condition can occur under the expected normal operating environment in vehicle communication where moving vehicles enters and exists other vehicles' communication range. 
This point implies that IEEE 802.11 OCB mode-based systems such as ETSI ITS-G5 and DSRC are suitable only for strictly broadcast-based communication in the V2X environment. 

Another issue of with 802.11 OCB mode is the lack of reliability and performance in V2X environment. 
A formal analysis by Ma et al.\ \cite{ma2009performance} shows that 802.11 OCB based system is not able to guarantee high reliability due to potential frame collisions and severe channel fading condition. 
The exponential back-off mechanism used in 802.11 to address frame collision and degraded radio condition has negative performance implications with increasing traffic load. 
This problem is pronounced further in dynamically moving vehicle environment. 
In addition, hidden terminal problem is more severe in broadcast than that in unicast. 
In other words, both broadcast and unicast modes have issues in 802.11-based systems in V2X environment.
This is in contrast to the conventional WLAN environment where wireless devices are likely more static compared to vehicles moving at high speed. 
On the contrary, in mobile systems (e.g.\ 4G LTE) radio resource is managed by the network. 
While various cellular protocols use time slotting or time sync to prevent the above-mentioned performance issue, latency comes however at a price. 

\subsection{ISO standard}
\label{sec:v2xTechnologies_ss4}
International Standard Organization (ISO) specifies several vehicle-related standards. 
ISO~26262~\cite{iso26262} defines functional safety of electrical and electronic devices for the automotive industry. It adapts the International Electrotechnical Commission (IEC) 61508 standard, a functional safety standard that defines safety life cycle of electronic systems and products for all industries. It is a risk-based safety standard; vehicles assess the risk of possible hazardous situations and mitigate their impacts to avoid systematic failure of vehicles. 
It was first published in 2011 and was revised in 2018~\cite{iso26262} in which cybersecurity aspects are added in a limited scope by covering only the interface from functional safety to cybersecurity~\cite{schmittner2018status}.

ISO/SAE~21434~\cite{isosae21434} specifies cybersecurity standard for road
vehicles. It started in 2016 as a joint work of ISO and SAE. It is based on
SAE~J2735~\cite{saeJ2735}; ISO/SAE~21434 defines a process and minimum
criteria for cybersecurity engineering through all phases of product life
cycle to prevent cyberattack on vehicles~\cite{macher2020iso}.
By complying to this standard, the whole automotive industry follows the
uniform cybersecurity development process through the vehicle
development life cycle. An analysis by Macher et al.~\cite{macher2020iso} indicates
that ISO/SAE~21434 leaves a gap as it stays at an abstract level and is not
intended to provide answer to specific implementation details, methods,
guidelines, or best practice, and does not present a
``silver-bullet'' per se. In addition, cybersecurity for autonomous vehicles and non-vehicles such as RSU are outside the scope of this standard. 

ISO~39001~\cite{iso39001} is a management system standard for Road Traffic Safety (RTS). It was first published in 2012. Its goal is to improve organizations' traffic safety, and it is targetted for organizations that have a process to improve traffic safety. Organizations that adhere to ISO~39001 can obtain certification of compliance.

% !TEX root = v2x_survey_main.tex
% !TeX spellcheck = en-GB

% ========  Section 4  ==========

\section{Security Management System Overview}
\label{sec:systemOverview}
In this section, we first revisit the security related standards in both
the EU and the US. Then we review the security management system defined in
these specifications. Both the EU and the US systems are based on a Public
Key Infrastructure (PKI) \cite{adams2003PKI}\@. These ITS security
management systems particularly focus on how to manage the certificates for
the ITS-Stations (ITS-S). This section highlights the similarities and 
differences between security management systems in the EU and the US as a baseline for the 
discussion in the following sections.

\subsection{ETSI C-ITS and IEEE 1609.2}
\label{sec:systemOverview_ss1}
\begin{figure}[h]
  \centering
  \includegraphics[width=0.6\linewidth]{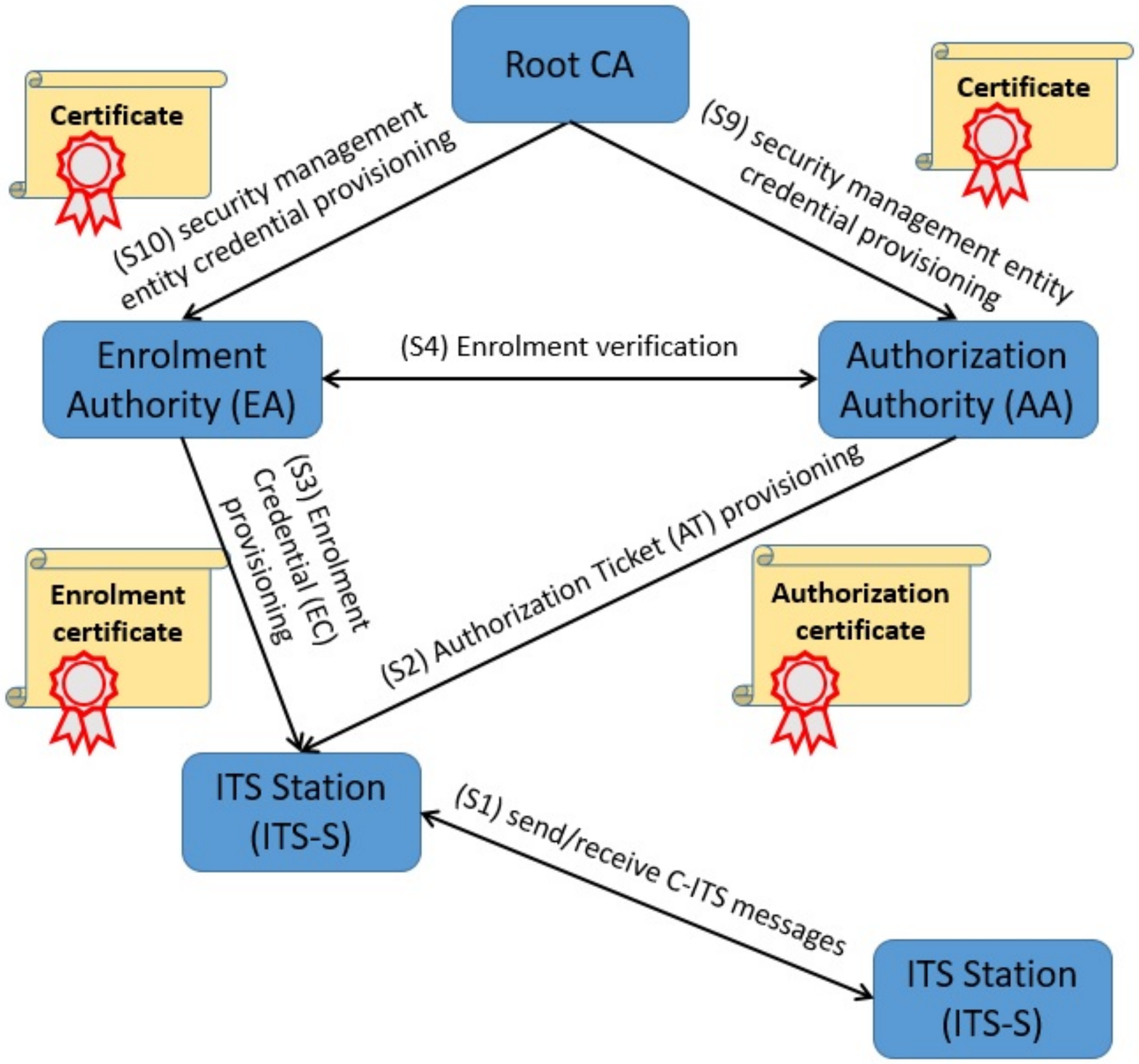}
  \caption{Overview of Security Management System}
  \label{fig:OverviewSecMgmtSystem}
\end{figure}

The most notable security-related ITS specifications are listed in Table~\ref{tab:ITSSecspecs}. 
The certificate management system adopted in the US is specified in IEEE~1609.2~\cite{intelligent2016ieee16092}. It is called Security Credential Management System (SCMS)\@. A comprehensive and detailed description of the SCMS is found in Brecht et al.~\cite{brecht2018security}.
The ETSI ITS standard covers a range of aspects across separate documents. 
Among them, ETSI TS 102~940~\cite{etsi2021102940} defines the overall architecture of C-ITS security management system. 
Figure~\ref{fig:OverviewSecMgmtSystem} illustrates the overall security management system and the relationship among the entities within the system, including the definition of reference points (e.g. S1, S2), following~\cite{etsi2021102940}. Key entities include:
\begin{itemize}
\item Root CA (RCA) –- This entity is the root of trust of the entire ITS certificate management system. One or more RCA issues certificates to the EA and AA\@.
\item Enrollment Authority (EA) –- This entity accepts enrollment requests from the ITS-S and issues enrollment credentials that are used by the ITS-S to contact the AA\@.
\item Authentication Authority (AA) –- This entity verifies the successful enrollment based on the enrollment credential issued by the EA, and issues one or more Authorization Certificates, which is also referred to as an Authorization Ticket (AT)\@. The AT is equivalent to a pseudonym certificate in general term.
\item ITS Station (ITS-S) –- This entity is the end device and the user of ATs issued by the AA. It includes multiple types of devices such as an On-Board Unit (OBU) in the vehicle, a Road Side Unit (RSU), and other types of devices that are engaged in the V2X communication.
\end{itemize}

\begin{table}[h]
	\footnotesize
	\vspace{-0.3cm}
	\caption{US AND EU ITS SPECIFICATIONS (SECURITY SPECIFIC)}
	\vspace{-0.3cm}
	\label{tab:ITSSecspecs}
	\begin{tabular}{llll}
	\noalign{\hrule height 1pt}
	Spec \#   & Title  & Latest Version  & Ref.  \\ 
	\noalign{\hrule height 1pt}
	\multicolumn{4}{c}{US Specification (IEEE)} \\ \hline
	IEEE 1609.2 & \begin{tabular}[c]{@{}l@{}}IEEE Standard for Wireless Access in Vehicular Environments—\\ Security Services for Applications and Management Messages\end{tabular} & 2016, January & \cite{intelligent2016ieee16092} \\ 
	\noalign{\hrule height 1pt}
	\multicolumn{4}{c}{EU Specifications (ETSI)} \\ \hline
	TS 102 731   & \begin{tabular}[c]{@{}l@{}}Intelligent Transport System (ITS); Security; Security Services and \\ Architecture\end{tabular}                                                       & V.1.1.1, 2010-09 & \cite{etsi2010102731} \\ \hline
	TS 102 940   & \begin{tabular}[c]{@{}l@{}}Intelligent Transport System (ITS); Security; ITS Communications \\ security architecture and security management\end{tabular}                         & V.2.1.1, 2021-07 & \cite{etsi2021102940} \\ \hline
	TS 102 941   & Intelligent Transport System (ITS); Trust and Privacy Management                                                                                                                     & V.2.1.1, 2021-10 & \cite{etsi2021102941} \\ \hline
	TS 102 942   & \begin{tabular}[c]{@{}l@{}}Intelligent Transport Systems (ITS); Security; Access Control \\ Technical Specification\end{tabular}                                                  & V.1.1.1, 2012-06 & \cite{etsi2012ts102942} \\ \hline
	TS 102 943   & Intelligent Transport Systems (ITS); Security; Confidentiality services                                                                                                              & V.1.1.1, 2012-06 & \cite{etsi2012ts102943} \\ \hline
	TS 103 097   & \begin{tabular}[c]{@{}l@{}}Intelligent Transport Systems (ITS); Security; Security header \\ and certificate formats\end{tabular}                                                 & V.2.1.1, 2021-10 & \cite{etsi2021103097} \\ 
	\noalign{\hrule height 1pt}
	\end{tabular}
\end{table}

\subsection{Differences between ETSI C-ITS and IEEE 1609.2}
\label{sec:systemOverview_ss3}
In this section, we describe differences in the security management systems between ETSI~ITS and IEEE~1609.2~\cite{intelligent2016ieee16092} (SCMS~\cite{brecht2018security}).

\subsubsection{Architectural Principles of Security Management}
\label{sec:systemOverview_sss3_1}
The SCMS architecture~\cite{brecht2018security} ensures strong privacy protection of vehicle owners by enforcing strict separation of vehicle information and the network entities under different organizations. This architecture allows no single entity in the security management system to have access to information sufficient to identify a vehicle: identifying a vehicle in SCMS requires multiple management entities under different organizations to collude. To achieve this goal, this architecture includes purpose-specific entities such as Registration Authority (RA) and a pair of Linkage Authorities (LAs). This level of functional and ownership separation is beyond the extent of ETSI ITS certificate management system~\cite{etsi2021102940}. To achieve the similar level of functional separation in ETSI ITS\@ as in the SCMS, extra design and implementation steps are required beyond the scope of the ETSI ITS standard.

\subsubsection{Revocation of ITS-S Certificates}
\label{sec:systemOverview_sss3_2}
The SCMS supports \textit{active revocation} of pseudonym certificates. \textit{Active revocation} means that the certificate management system revokes the pseudonym certificates of a vehicle by issuing a Certificate Revocation List (CRL). The SCMS defines separate entities dedicated for this task, such as Misbehaving Authority (MA) and a pair of LAs. When a vehicle observes misbehavior of another vehicle, the former reports to the MA by including the latter vehicle's \textit{linkage value} which is included in its pseudonym certificate. Then the MA determines whether misbehavior exists or not by correlating multiple reports and verifying the alleged misbehavior. 

Upon confirming the positive misbehavior, the MA resolves the reported \textit{linkage value} to two \textit{linkage seeds} by contacting the Pseudonym CA (PCA), the RA, and the LAs. The MA contacts these three entities in a serial manner in this process as each entity has only limited information that collectively triggers the LAs to retrieve the correct \textit{linkage seeds} of the to-be-revoked vehicle.

After the \textit{linkage seeds} are obtained from the LAs, the MA creates
an entry in the CRL by including the tuple of (\textit{linkage seeds},
current time period, and the number of simultaneously active pseudonyms)\@.
As vehicles receive the CRL, they use this tuple and reconstruct a set of
\textit{linkage values} that correspond to all pseudonym certificates of
the revoked vehicle. This way, vehicles can identify revoked certificates
by comparing the \textit{linkage value} within the certificate in the
received messages against the values reconstructed out of the CRL. 
This way, the number of simultaneously valid certificates or future certificates preloaded to vehicles does not impact the CRL size (cf. Sec.~\ref{sec:issueUseofCertificates_ss3}).  
We refer the reader to~\cite{brecht2018security} for further details.

ETSI~ITS does not define \textit{active revocation} of certificates. Instead, it solely relies on a \textit{passive revocation} mechanism. \textit{Passive revocation} is accomplished by denying further allocation of certificates to a vehicle when the system determines that the vehicle needs to be revoked. This rejection occurs at the time when the vehicle attempts to obtain additional pseudonyms from the certificate management system.

\subsubsection{Certificate Issuing and Usage Schemes}
\label{sec:systemOverview_sss3_3}
Annex 2 in the 5GCAR~D4.1 document~\cite{gallo20185gcard41} describes the security architecture of the US and the EU certificate management systems. In the US system, the certificate management system (more specifically the RA and PCA collectively) generates three-years worth of certificates and preload them to a vehicle, containing 20 pseudonym certificates per week to a vehicle~\cite{brecht2018security,gallo20185gcard41}. This pseudonym size is derived from C2C-CC recommendation~\cite{bissmeyer2011generic} as stated in SCMS~\cite{brecht2018security}.

An ETSI report on pseudonym change management in TR 103 415~\cite{etsi2018103415} lists six different types of pseudonym management schemes. Given the nature of a pre-standard report, it does not yet specify exact details in this area. It indicates that the pseudonym pool size in referenced schemes varies between 10 to 100 depending on the scheme. The underlying mechanism is that a vehicle cycles through a set of certificates for a fixed duration of a week. The smallest size of 10 is the proposal from SCOOP@F project~\cite{scoopUrl} and the largest pool size of 100 is the recommendation by EC's security policy and governance frame work~\cite{EC2017securitypolicy} and certificate policy~\cite{EC2018certificatepolicy}. One exception is Issue First Activate Later (IFAL) scheme~\cite{verheul2019ifal} which strictly uses only one pseudonym at a time without reuse.

Neither the aforementioned 5GCAR D4.1~document~\cite{gallo20185gcard41},
SCMS~\cite{brecht2018security}, nor the ETSI ITS
specifications~\cite{etsi2021102940,etsi2021102941} explicitly specify the
change period of one certificate from another within a one-week period.
This period influences a vehicle’s vulnerability to tracking and identification. Multiple schemes may be employed for this purpose; this period may be static for all vehicle types under all circumstances, or may vary depend on vehicle types and specific circumstances.

% !TEX root = v2x_survey_main.tex
% !TeX spellcheck = en-GB

% ========  Section 5  ==========
\section{Attack Types and Threats} 
\label{sec:threatModel}
Before we discuss individual security and privacy-related issues in the subsequent sections, we first identify attack types and their resulting threats.  
Table~\ref{tab:attack_types} captures our view on this point. 
This table is not intended to be exhaustive; 
in fact, there may be new types of attacks we are not aware of today but become possible in the future. However, it is important to be cognizant to this important aspect.

\begin{table}[h]
	\footnotesize
	\vspace{-0.3cm}
	\caption{Attack Types and Threats}
	\vspace{-0.3cm}
	\label{tab:attack_types}
	\begin{tabular}{lll}
		\noalign{\hrule height 1pt}
		Attack types   & Attack Description  & Resulting Threats \\ 
		\noalign{\hrule height 1pt}
		\begin{tabular}[c]{@{}l@{}} \textit{Passive} vs. \\ \textit{Active} \end{tabular} 
			& \begin{tabular}[c]{@{}l@{}}
				\textit{Passive}: monitor communication channels and \\
				obtain information from it. \\
				\textit{Active}: send disruptive messages to the \\
				communication channels or breaks in a system. 
			\end{tabular} 
			& \begin{tabular}[c]{@{}l@{}}
				\textit{Passive}: collected data can identify or trace \\
				vehicles.\\
				\textit{Active}: disruptive messages may cause accidents, \\
				or intrude the system resulting in information \\
				loss or system malfunction.
			\end{tabular} \\ \hline
		\begin{tabular}[c]{@{}l@{}} \textit{Local} vs. \\ \textit{Remote} scope\end{tabular} 
			& \begin{tabular}[c]{@{}l@{}}
				\textit{Local}: passively monitor or actively send \\
				disruptive messages at a specific location to its \\
				immediate area. \\
				\textit{Remote}: passively monitor or actively send \\
				disruptive messages to/from one or more remote \\
				locations.
			\end{tabular} 
			& \begin{tabular}[c]{@{}l@{}}
				\textit{Local}: the scope and impact of data collection \\
				and disruptive messages is limited to a specific \\
				area only. \\
				\textit{Remote}: the scope and impact of data collection \\
				and disruptive messages spans farther to wider \\
				areas.
			\end{tabular} \\ \hline
		\begin{tabular}[c]{@{}l@{}} \textit{Local} vs. \\ \textit{Global view} \end{tabular} 
			& \begin{tabular}[c]{@{}l@{}}
				\textit{Local view}: data collection in a limited scope, e.g. \\
				using a single or small number of devices in a \\
				limited area. \\
				\textit{Global view}: data collection and aggregation from \\
				large number of devices in a wide area. 
			\end{tabular} 
			& \begin{tabular}[c]{@{}l@{}}
				\textit{Local view}: obtains traffic flow or pattern within \\
				a limited area.\\
				\textit{Global view}: obtains traffic flow or pattern in a \\
				wide area, such as entire country. 
			\end{tabular} \\ \hline
		\begin{tabular}[c]{@{}l@{}} \textit{Insider} vs. \\ \textit{Outsider} \end{tabular} 
			& \begin{tabular}[c]{@{}l@{}}
				\textit{Insider}: an employee of RSU infrastructure \\
				system steals data or disrupts its operation. \\
				\textit{Outsider}: a hacker builds a device that sends \\
				messages as a fake vehicle, or breaks in a system. \\ 
			\end{tabular} 
			& \begin{tabular}[c]{@{}l@{}}
				\textit{Insider}: loss or leak of data that are otherwise \\
				available only to insiders. \\
				\textit{Outsider}: disruptive messages from a fake device \\
				cause negative consequence such as accidents. 
			\end{tabular} \\ \hline
		\begin{tabular}[c]{@{}l@{}} \textit{Individual} vs. \\ \textit{Organized} \end{tabular} 
			& \begin{tabular}[c]{@{}l@{}}
				\textit{Individual}: a motivated hacker with limited budget \\
				and materials with an intent to disrupt \\
				communication. \\
				\textit{Organized}: an organized group (e.g. nation state) \\
				with unlimited resources with large budgets, \\
				facilities, and materials with intentions to disrupt \\
				an enemy nation.
			\end{tabular} 
			& \begin{tabular}[c]{@{}l@{}}
				\textit{Individual}: limited impact relative to the effect a \\
				single individual can cause, e.g. a small number \\
				of fake devices.\\
				\textit{Organized}: more organized and larger-scale \\
				attacks possible using a dedicated infrastructure.
			\end{tabular} \\ 
		\noalign{\hrule height 1pt}
	\end{tabular}
\end{table}
Different types of adversaries have different motivations and goals. An individual hacker may have fun out of disrupting society, and would likely be satisfied to see the resulting chaos in reality. 
On the other hand, a large organized crime group may aim to disrupt peace in an enemy nation with an intention to cause chaos, physical and material damages, and panic. 
In both cases, adversaries' motivations and goals are related to the aspects of the CIA triad~\cite{covert2020towards}. 
\begin{itemize}
	\item Reduced confidentiality: as a result of compromising privacy of vehicle owners by tracking and identifying vehicles.
	\item Reduced system integrity: by making it untrustworthy, e.g. by introducing fake vehicles injecting false messages.
	\item Reduced system availability: by disrupting the objective to promote road safety, such as by creating denial-of-service (DoS) situation.
\end{itemize}

% !TEX root = ./v2x_survey_main.tex
% !TeX spellcheck = en-GB

% ========  Section 5  ==========

\section{Security Issue: Privacy Protection}
\label{sec:issuePrivacyProtection}
One of the goals of the security management system in V2X communication is privacy protection, i.e.\ protecting the vehicle owners’ privacy as required by relevant regulations. In the EU, these are GDPR \cite{regulation2016regulation}, Network and Information System (NIS) Directive \cite{directive2016directive},  the Cybersecurity Act~\cite{cybersecurityact2019}, and the ePrivacy Directive \cite{parliament2002directive}. Thus, user privacy protection is a requirement for V2X communications.
Privacy protection technologies aim to prevent attacks or to confuse
attackers who attempt to track vehicles by intercepting
communications or tracing V2X interactions.
A range of privacy protection strategies have already been developed and
partially standardized. It is important to ensure that these
privacy-preserving approaches do not impede safety functions which rely on
vehicular communications.

A key strategy to achieve privacy protection is to rely on periodically
changing pseudonyms for all communication involving vehicles. To
provide pseudonyms to vehicles, certificate management systems
such as SCMS \cite{brecht2018security} have been designed for the US 
based on PKI~\cite{adams2003PKI} (cf.\ Sec.~\ref{sec:systemOverview_sss3_3}). 
The EU security architecture is based on the same approach. 
Privacy protection includes the following concepts as defined by Pfitzmann and Hansen~\cite{pfitzmann2010terminology}:
\begin{itemize}
\item Anonymity: \textit{``Anonymity of a subject means that the subject is not identifiable within a set of subjects, the anonymity set.''}
\item Pseudonymity: \textit{``Pseudonymity is the use of pseudonyms as identifiers''}
\item Unlinkability: \textit{``Unlinkability of two or more items of interest (IOIs, e.g., subjects, messages, actions, \ldots) from an attacker’s perspective means that within the system (comprising these and possibly other items), the attacker cannot sufficiently distinguish whether these IOIs are related or not.''}
\item Unobservability: \textit{``Unobservability of an item of interest (IOI) means (i) undetectability of the IOI against all subjects uninvolved in it and (ii) anonymity of the subject(s) involved in the IOI even against the other subject(s) involved in that IOI.''}
\end{itemize}

Protecting the vehicle owners’ privacy is a meaningful goal. However, there are several scenarios worth considering for the applicability of privacy protection. They appear to directly conflict with the privacy protection requirement.

\subsection{Privacy and Vehicle Operation}
\label{sec:issuePrivacyProtection_ss0}
Vehicles with no malicious intent store certificates from other vehicles as a part of their normal operation. How long vehicles store these certificates have implications on privacy of vehicle owners. 
Vehicles store other vehicles' certificates for several reasons. First, verification of sender's authenticity requires verifying the entire certificate chain from up to the RCA. However, messages may not necessarily contain all certificates in the chain. Therefore, once a vehicle obtains the entire certificate chain for a given AT, it needs to keep this set for future references.  
Second, some message types (i.e. CAM, cf. Sec.~\ref{sec:issuePrivacyProtection_ss3}) do not always contain an AT in every message. 
This implies that receiving vehicles need to store a copy of this AT to verify the message integrity as long as the transmitting vehicle uses the said AT. 
Therefore, the difference between malicious and benign devices are subtle: presence or absence of malicious intent to use collected information.  
In this sense, privacy protection implicitly includes protection against benign vehicles also. 
We consider this is a fundamental constraint of the certificate-based message verification in broadcast-mode. 
This point raises a question: when does a benign vehicle deletes \textit{old and stale} ATs that has been stored in it but no longer used due to, either the AT-owner vehicles changed their ATs, or they moved out of the communication range from the vehicle. 
Another question is a requirement to retain received ATs for forensics purposes, such as investigating accidents.  
If such requirement exists, it may vary in countries or jurisdictions, thus likely not a one-size-fits-all answer. In this respect, these questions are implementation-dependent matter. 
Differences in these aspects likely influence the privacy of vehicle owners as how long privacy-related information is stored in other vehicles.  
The ETSI standard does not address either of these questions, leaving as an open issue in the standard.

\subsection{Applicability of Privacy 1: Vehicle Types and Usages}
\label{sec:issuePrivacyProtection_ss1}
There are multiple types of vehicles on the road, and the privacy requirements are not likely to be applied to all of them uniformly. ETSI TS 102~941~\cite{etsi2021102941} specifies the privacy requirements for ITS\@. However, it does not consider the option to apply different privacy measures depending on the vehicle type.

First, not all vehicles are privately owned. Some special vehicles and non-passenger vehicles are owned by a company rather than by an individual. Trucks, delivery vans, and taxis fall into this category. These vehicles quite often have a company name or a logo written on the vehicle’s body. Even a privately-owned vehicles may have a name, an address, an email address, and a telephone number written on the vehicle’s body if its owner has a private business to advertise. In this sense, many vehicles voluntarily forfeit the privacy information. Although drivers or vehicle owners wish to protect their privacy, it is an imbalance between volunteering visible information and a need to prevent \textit{remote} tracking. 
Second, some special vehicles, such as police cars, ambulances, and public transportation (e.g.\ buses), belong to various levels of government or public entities rather than an individual. Hence, in this case, different level of privacy protection may apply for these vehicles while considering their minimum level of protection against tracking.

According to ETSI TS 102~941~\cite{etsi2021102941}, OEMs are expected to assign a canonical permanent vehicle ID to each vehicle at the time of manufacturing. This permanent ID is what the privacy protection requirement intends to protect by using pseudonyms instead. However, OEMs do not know for what purpose any given vehicle will be used at the time of manufacturing. For example, for the exact same type vehicles, one of them may be used by an individual owner; another may be used by a business to which no particular individual is associated with.
The former case requires more strict privacy protection than the latter. 
Therefore, it is likely that we need to rely on other mechanisms to determine what level of privacy a given vehicle requires. This type of consideration is not given in the ETSI ITS specifications.

Another related aspect is the change of vehicle ownership. When a vehicle is sold in the second market, the privacy-related information of the previous owner stored in the vehicle needs to be erased. 
This includes data such as unused ATs, navigation history, and Bluetooth-paired smartphone.  
The sales process of second-hand vehicles needs to include the necessary procedure to erase these data. 
As vehicles can be sold by owners rather than by auto dealers, it should be a simple process to trigger it through vehicle's user interface (UI). 

\subsection{Applicability of Privacy 2: Non-Vehicle ITS-S}
\label{sec:issuePrivacyProtection_ss2}
There are different types of ITS-S, most prominently the vehicles, and the road-side RSU\@.
ETSI ITS TS 102~940~\cite{etsi2021102940} describes the overall certificate management system. However; the standard \cite{etsi2021102940} only considers a vehicle-centric view of the system; it has no description on the certificate management for RSUs as another type of ITS-S.

As discussed in Sec.~\ref{sec:issuePrivacyProtection_ss1}, privacy requirements are intended to protect the privacy of vehicle owners as
private individuals. However, RSUs do not have any private owner or person
associated with it. Therefore, it follows that RSUs do not need pseudonyms.  The fact that privacy requirements may
depend on the ITS-S type is not considered in the ETSI ITS specifications.
How pseudonym certificates are managed for RSUs, or whether
they are necessary for RSU at all, is not specified. See the related discussion in Sec.~\ref{sec:issueUseofCertificates_ss5}.

\subsection{Privacy and Cooperative Awareness}
\label{sec:issuePrivacyProtection_ss3}
Cooperative Awareness Messages (CAM), as specified in ETSI TS 302.637-2~\cite{etsi20193026372}, share basic attributes of the vehicles on the road. These attributes include the vehicle’s position, speed, direction, acceleration, vehicle length and width, vehicle type, etc. Vehicles on the road periodically transmit CAM messages to mutually establish and maintain situational awareness in the vicinity. 
At the same time, vehicles change their pseudonyms periodically to preserve privacy and prevent tracking, as specified by EN 302 636-6-1~\cite{etsi201430263661}. They also need to change their MAC address at the same time (and its IPv6 address in case of IP-based communication). This simultaneous changes of pseudonym, MAC and IPv6 address prevents adversaries and other vehicles from linking consecutive pseudonyms by either the MAC or IPv6 address. 

Despite this privacy-protection scheme, broadcasting vehicle information in CAM can reveal sufficient information for receiving vehicles to identify the transmitting vehicle. 
For example, if there are only a few vehicles in the distinctive positions relative to the receiving vehicle (e.g.\ one in front and another behind), then the receiving vehicle can correlate the CAM messages with the vehicle by comparing the received position information against its own position. 
In another example, if a received CAM message indicates the transmitting vehicle is 15-meter long and there is only one 18-wheeler truck in the vicinity, it is also trivial to correlate this message to that vehicle. 
In fact, study by Escher et al.~\cite{escher2021well} found that additional information such as vehicle size \textit{``enormously improves''} the pseudonym linkage, allowing tracking of up to 80\% of vehicles. 
Further, the number of vehicles in the vicinity plays an important factor. This study concludes that the location privacy will decrease despite the change of pseudonyms. 

Further, if vehicles collect and store pseudonym changes from vehicles based on relative position information, share and collaborate this information in a wider-scale, for example by uploading it in a cloud storage, real-time tracking of vehicle movement, such as city-wide level or even larger scale, would become possible. 
This way, vehicle tracking may become similar to what already exists in real-time flight and ship tracking maps such as in~\cite{flightradar24Url,marinetrafficUrl}. 

In this way, simply transmitting CAM messages including position information and other attributes already helps a receiving vehicle to identify the transmitting vehicle, which leads to a possible compromise of the privacy of the vehicle owner. ETSI TS 102~940~\cite{etsi2021102940} Clause~4.3.1.3 states: \textit{``\ldots it is necessary to ensure that the data cannot be linked to any individual so that no personally identifying information is leaked by the CAM service.''} If we take this requirement text literally, the CAM message itself does not reveal the personally identifying information. However, the content in the CAM message provides information that certainly helps to violate the principle of privacy protection by linking the transmitted message and the vehicle that transmitted it.

One possible approach to mitigate this situation is to enforce strict one-time use of pseudonym. However, it still does not guarantee the unlinkability property if series of pseudonym changes are observed by and shared with multiple vehicles through the cloud storage.  
This approach also increases the required number of pseudonyms per vehicle; it will stress the management system to generate and deliver pseudonyms and vehicles to store them. The pseudonym change scheme as specified in standards requires further research.

\subsection{Privacy vs. Road Safety}
\label{sec:issuePrivacyProtection_ss4}
\begin{figure}[h]
  \centering
  \vspace{-0.3cm}
  \includegraphics[width=0.35\linewidth]{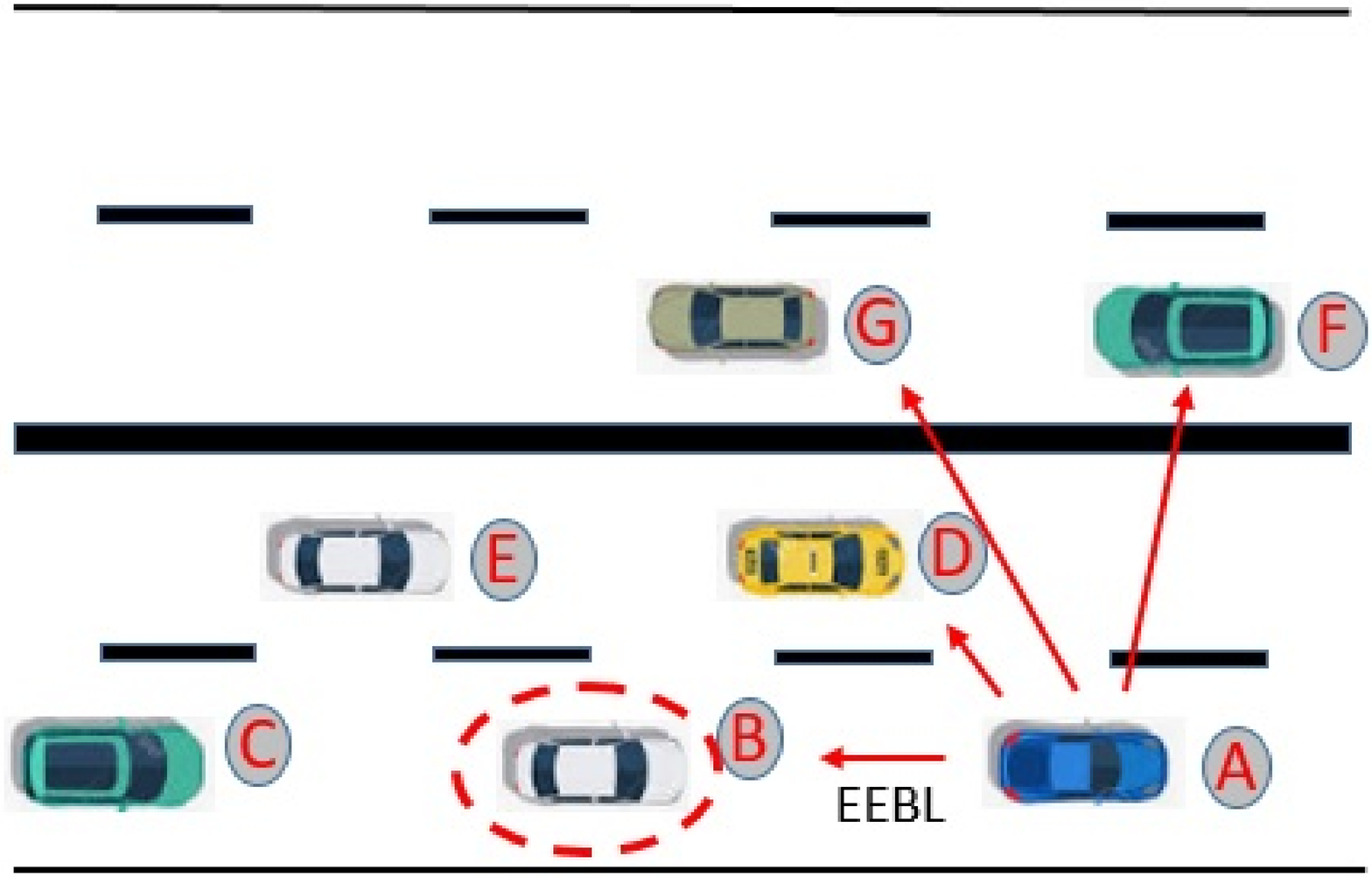}
  \vspace{-0.3cm}
  \caption{DENM and Privacy Requirement}
  \label{fig:DENMandPrivacyRequirement}
  \vspace{-0.5cm}
\end{figure}

Privacy protection is at odds with road safety. Consider a simple scenario where there are multiple vehicles in a multi-lane road as shown in Figure~\ref{fig:DENMandPrivacyRequirement}. Vehicle~A suddenly applied a brake which triggers the broadcast of an emergency electronic brake light (EEBL) message to the surrounding vehicles. The EEBL message is one of the De-centralized Environmental Notification Messages (DENM) as defined in ETSI TS 302.637-3~\cite{etsi20193026373}. In this case, Vehicle~B, which is directly behind Vehicle~A, needs to brake immediately to avoid an imminent collision. This is an essential requirement to reduce the number of road accidents. However, due to the privacy requirements, TS 102 940~\cite{etsi2021102940} states that the pseudonym used in DENM must be different (unlinkable) from the one used in CAM. This way, the message origin of the EEBL is kept anonymous and unobservable, meaning that Vehicle~B is neither expected to know which vehicle originated this EEBL message, nor can it determine whether a specific vehicle (e.g.\ Vehicle~A) transmitted this message or not. Therefore, it follows that all vehicles not only cannot determine whether to apply its brake or not, but also making a wrong decision can make the situation even more dangerous, e.g. Vehicle~D or G applying a sudden braking for no apparent reason. At the same time, all DENM messages, including EEBL, include position information of the transmitting vehicle according to the specification in 
ETSI TS 302~637-03~\cite{etsi20193026373}. This is a contradiction to the privacy requirement (cf.\ Sec.~\ref{sec:issuePrivacyProtection_ss3}). 

This simple example illustrated above indicates that the privacy protection requirement is directly at odds with road safety. This leads to several issues:
\begin{itemize}
\item The privacy protection requirement hinders other vehicles to identify where a message comes from.
\item Lack of this knowledge hinders surrounding vehicles to make right decisions to prevent an accident.
\end{itemize}
The analysis by Chator and Green \cite{chator2018don} also identified these points in the security requirement in the ETSI ITS specifications; we agree with this analysis. Based on the discussion above, we conclude that the privacy requirement in ETSI ITS specifications needs to be reconsidered.

\subsection{Privacy and Use of Unicast}
\label{sec:issuePrivacyProtection_ss5}
Some use cases in the EU projects \cite{Concordaurl,5gcarurl,5gcarmenurl,5gcrocourl,5gmobixurl} are based on unicast communication between two endpoints. Some examples include remote driving or tele-operated driving use cases in the 5GCAR \cite{fernandez20195gcard21} and 5G CroCo project \cite{perraud20195gcrocod21}. Unicast communication has an advantage over broadcast in that the former can use confidentiality protection through encryption. This prevents passive observers from identifying the information transmitted by a vehicle. Although there are methods to apply confidentiality protection in broadcast, such as those proposed in \cite{du2005id,garay2000long,sakai2007identity,halevy2002lsd}, the ITS specification (TS 102~943~\cite{etsi2012ts102943}) does not require it to broadcast-based services such as CAM and DENM\@.

Despite the use of confidentiality protection, unicast communication conflicts with the privacy requirement. This applies to both internal and external threats. The internal threat refers to the leakage of private information between two endpoints of the communication; the external threat refers to the leakage of private information to another entity outside of these two endpoints.

First, we discuss the internal threat. If two entities establish a unicast communication, then by definition, both of them uniquely identify the other endpoint with the pseudonym, IP or MAC address of the other endpoint. In this case, applying a periodic pseudonym change is meaningful to protect the communication from eavesdroppers (i.e.\ protection against external threat). It makes it difficult for them to track unicast communication over a period longer than the pseudonym change cycle. However, it also means that the \textit{linkability} of the old and the new pseudonym is voluntarily shared with the peer endpoint of the unicast communication. The peer vehicle can keep this information even after the unicast communication ends until the time that the vehicle changes its pseudonym again later. In addition, changing pseudonyms in the middle of unicast communication requires an explicit coordination between the two endpoints to maintain the communication. If this procedure fails for any reason, it can result in a negative consequence, such as a dropped communication. 

The change of pseudonyms in unicast communication requires special handling, similar to Network Mobility (NEMO) in RFC~3963~\cite{devarapalliietf3963}. It operates at the IP layer as described in ETSI EN 302~636-6-1~\cite{etsi201430263661}. Therefore, an additional mechanism is needed to handle link layer address changes. 3GPP TR 33.836~\cite{3gpp2019tr33836} defines several variations of such explicit link layer address change notification. These schemes are workable solutions. However, as stated earlier, they fundamentally violate the \textit{unlinkability} requirement between two endpoints in exchange for maintaining the unicast communication.

Second, from an external threat perspective, pseudonym changes in the ongoing unicast communication require confidentiality protection. Otherwise, it would be trivial for eavesdroppers to intercept the pseudonym change messages sent in clear and correlate the pseudonyms, hence violating the \textit{unlinkability} requirement.

Another possible approach is simply avoiding the pseudonym change until the unicast communication is completed. 
This is especially meaningful if confidentiality protection is not applied. In this case, both vehicles can change their pseudonyms as soon as the unicast communication ends. Doing so ensures that the \textit{unlinkability} principle is maintained. This approach is also beneficial as it eliminates potential failure of pseudonym change and communication loss in the middle of unicast communication. Further consideration is needed to ensure reliable unicast communication while satisfying the privacy requirement.

% !TEX root = v2x_survey_main.tex
% !TeX spellcheck = en-GB

% ========  Section 6  ==========

\section{Security Issue: Use of Certificates}
\label{sec:issueUseofCertificates}
Both the EU and the US systems are based on Public Key Infrastructure (PKI)~\cite{adams2003PKI} and use of certificates to manage the pseudonym usage. 
In this section, we focus specifically on aspects related to their usage. We will cover the actual issues and lack of definitions in the two competing standards (ITS-G5 and C-ITS) impacting certificate usage, renewal and revocation concerning vehicles, roadside infrastructures as well as pedestrians.
\subsection{Certificate-Based Message Verification}
\label{sec:issueUseofCertificates_ss1}
In V2X communication, a certificate is attached to a message and receivers of this message use the public key in the certificate to verify the message's authenticity. 
The PKI in charge of managing certificates~\cite{adams2003PKI} requires that the receiving entity should be able to verify the certificate chain up to the root CA in order to verify the message authenticity. This is the underlying assumption of using certificates in real-time communication.

Contrary to this assumption, the real-time verification of the certificate may not be possible under all circumstances. It is especially the case in a dynamically changing communication environment involving moving vehicles in open space. 
For example, CAM messages do not always contain a certificate. CAM messages contain a certificate at least once a second. However, when they are sent more frequently, a \textit{digest} (the least significant 8 octets of a hash output of a certificate) is added to replace a certificate. This use of \textit{digest} enables a compact representation of a certificate without sending it in every CAM message. TS 103 097~\cite{etsi2021103097} states that, if a vehicle receives a CAM message with unknown digest or a received certificate is signed by an unknown AA, then the receiving vehicle needs to resolve this situation by requesting the missing certificate to the surrounding vehicles and wait for a response (\textit{inlineP2pcdRequest}). Only after this step, the vehicle can verify the validity of the received message. This additional message exchange causes delay in message verification in the order of several 100 milliseconds at least.
This situation contradicts with the underlying expectation to process vehicular communication in real-time under all circumstances. 
Due to these issues, message identification based on certificate verification as a mean to ensure authentication in V2X communications remains an open issue.

\subsection{Certificate Usage and Change Policy}
\label{sec:issueUseofCertificates_ss2}
As discussed in Sec.~\ref{sec:systemOverview_sss3_3}, the certificate management in the EU issues a set of ATs per week~\cite{gallo20185gcard41}. ATs used in the V2X communication are expected to be changed periodically in the order of minutes. This implies that a set of ATs are reused multiple times during a specific one-week period. However, the exact duration of one AT and the mechanism to select the next one is not specified according to ETSI TR 103~415~\cite{etsi2018103415}. If ETSI standard stops at the level of recommendations and leave details to implementations, there will likely be variations in terms of its effectiveness in preventing adversaries from predicting the next AT. In this sense, careful considerations are required as sub-optimal usage and change policy can lead to successful linking of the certificate to the vehicle. We also discussed the open issue of pseudonym usage and its implications on privacy protection in Sec.~\ref{sec:issuePrivacyProtection_ss3}. Thus, it is an area that needs further consideration.

\subsection{Certificate Reloading}
\label{sec:issueUseofCertificates_ss3}
Another area not specifically defined in the standard is the reloading  of
certificates. Under any circumstances, situations need to be avoided where
a vehicle runs out of valid certificates, preventing that vehicle from
sending messages altogether. Several research papers discuss certificate reloading schemes such as in \cite{qiu2019secure}. These schemes use an RSU as an entry point to communicate with the AA\@. In these schemes, the underlying assumption is that an RSU is available whenever and wherever it is needed. However, RSUs may not necessarily be ubiquitous. Therefore, a universally workable solution is necessary so that legitimate vehicles can obtain new set of certificates without relying on the presence of RSU.

One possible approach is to store sets of certificates beyond the immediate period. Storing 3-years worth of certificates up front, as described in SCMS~\cite{brecht2018security} and the 5GCAR D4.1~document \cite{gallo20185gcard41}, is one such approach. Such scheme, at least in theory, alleviates the reloading needs for the duration of three years. However, it comes at the expense of additional storage to hold this amount of certificates in the vehicle. Also, longer term storage of certificates complicates the system if they need to be revoked due to, for example, the vehicle being identified as malicious, or adversaries steal valid certificates from a legitimate vehicle and use them for malicious purposes. 
Another consideration is handling of unused certificates when the vehicle is deregistered, or when the ownership is transferred to someone else. 

Another possible approach is to reload certificates for multiple periods in the future, or request the next set of certificates well before the currently stored sets are exhausted -- analogous to refilling the gas tank well before it is empty. For example, a vehicle stores sets of $n$ consecutive weeks worth of certificates, and requests the next sets well before the end of the $n$-th week. This approach gives extra time in case network connection is not available at the first attempt to contact the AA. In this case, the vehicle can retry within the remaining time. This approach avoids potential exhaustion of certificates at the end of every one-week cycle, and does not require large storage capacity compared to storing 3-years worth of certificates. 
As of today, the challenge to propose a system ensuring a continuous flow of valid certificates while contextually relevant and memory efficient for V2X message verification remains open.

\subsection{Certificate Revocation}
\label{sec:issueUseofCertificates_ss4}
\subsubsection{Active Revocation}
\label{sec:issueUseofCertificates_sss4_1}
Active revocation of certificates is an area that is distinctively different between the US and the EU systems as previously discussed in Sec.~\ref{sec:systemOverview_sss3_2}. The US system based on IEEE 1609.2~\cite{intelligent2016ieee16092} (SCMS~\cite{brecht2018security}) supports active revocation of certificates while ETSI ITS does not. Active revocation involves two aspects: 1) detection, reporting, and validation of vehicle misbehaviour, and 2) generation and distribution of the CRL. As the first aspect relative to handling misbehaviours is discussed later in this paper, in this subsection we focus on the latter -- CRL generation and distribution.

Management of the CRL, including its generation and distribution, is already challenging in a conventional PKI-based system~\cite{slagell2006survey}. Notable difficulties are guaranteeing the distribution of the CRL in a timely manner and keeping up with the scale of the distribution itself. It is even more challenging with moving vehicles. Because vehicles are assigned with a set of certificates valid for a limited period as discussed in Sec.~\ref{sec:issueUseofCertificates_ss2}, the number of certificates per vehicle significantly impacts the CRL size. This situation requires a technique to aggregate current and future certificates that belong to a vehicle and represent them in a compact manner for efficient distribution. The situation becomes even more prominent if a large sets of certificates are preloaded to the vehicle ahead of time, as discussed in Sec.~\ref{sec:issueUseofCertificates_ss3}. 
We previously explained in Sec.~\ref{sec:systemOverview_sss3_2} that in order to address this issue, the SCMS describes the revocation scheme using \textit{linkage value}. It addresses one aspect of the distribution issues by reducing the amount of revocation-related information per vehicle. However, 
the CRL size is still a concern as the number of revoked vehicles increases over time. SCMS~\cite{brecht2018security} does not address this aspect; it is most likely left as a deployment-level matter as it depends on the size of the vehicle population under a certificate management system.

Other issues of the revocation process remain open, especially the ones related to efficiently distributing the CRL. 
The main question on this issue is how to ensure the latest CRL is made available to vehicles in a timely manner. Even if the latest most up-to-date CRL is generated correctly, if it is not delivered to vehicles that need it when they need it, it is of little value. 
Online Certificate Status Protocol (OCSP) specified in 
RFC~6960~\cite{santesson2013ietf6960} is an alternative approach to the use of CRL. However, it does not solve the issue as OCSP suffers from the same issue of absence of guaranteed connection with the network under all circumstances~\cite{samoshkinUrl}.

The other related issue of CRLs distribution is their size versus the vehicle’s storage capacity. Clearly, it is not realistic for a vehicle to store all revoked certificates of all vehicles. It is straightforward to consider that vehicles need to store CRL of vehicles that are \textit{relevant} to them. By \textit{relevant}, we mean related to vehicles that they may encounter on the road. There is no point of receiving and storing CRLs concerning vehicles that the receiving vehicle never comes across. However, how to determine the relevance is a matter of context for each individual vehicle. It depends on where a given vehicle drives, e.g.\ which country, region or province, highway or street, etc. Ideally, all vehicles that encounter a given revoked vehicle should be provided with the CRL containing the revoked certificates of that vehicle in question. If a vehicle misses a specific vehicle's certificates in its CRL and receives a message from this revoked vehicle, it does not know that it should reject all messages sent by this revoked vehicle. 
How to determine the relevance, or even the concept of CRL \textit{relevance}, is not defined in IEEE 1609.2~\cite{intelligent2016ieee16092} (SCMS~\cite{brecht2018security}). Thus, it most likely falls in the implementation-dependent area. It is certainly not a trivial problem to solve, making it a possible further research area.

Previous research has been conducted on the subject of CRL distribution in V2X context proposing various schemes to make it efficient. Some of these schemes include: 1) splitting CRL in small pieces~\cite{khodaei2020scalable,nowatkowski2010certificate}, 2) distribution through RSU~\cite{papadimitratos2008certificate}, 3) CRL dissemination in epidemic fashion~\cite{laberteaux2008security}, 4) use of Bloom Filter to reduce the CRL size~\cite{rigazzi2017optimized}. Although these works include novel approaches to increase efficiency of CRL distribution, they do not address the \textit{relevance} aspect we discussed.
 
Another issue related to CRL distribution is how to determine expired entries and when to remove them to prevent the CRL size from growing indefinitely over time. 
Even if a vehicle is already de-commissioned and thus is no longer on the road, it does not necessarily mean that these entries can be removed from the CRL as it is necessary to prevent the situation where adversaries can steal valid certificates and corresponding private keys from such vehicle to pose as a legitimate vehicle. Although many research papers focus on the CRL distribution, there is little attention to this area. 
This is not a trivial question as it closely relates to the preloading and reloading of certificates (cf. Sec.~\ref{sec:issueUseofCertificates_ss3}), i.e. the longer the perloading and reloading period, the longer the entry needs to remain in the CRL. This is also an open question that the IEEE 1609.2 standard~\cite{intelligent2016ieee16092} needs to address.

\subsubsection{Passive Revocation}
\label{sec:issueUseofCertificates_sss4_2}
As discussed in Sec.~\ref{sec:systemOverview_sss3_2}, the EU system does not require active revocation of certificates, thus relies solely on passive revocation. Passive revocation is a scheme based on a \textit{blocklist}. A malicious or misbehaving vehicle is blocklisted. Thus, when a vehicle sends a request to reload certificates next time, the certificate management system denies the request if the requesting vehicle is on the \textit{blocklist}.

One advantage of passive revocation is its simplicity as the certificate management system alleviates itself from the trouble of generating and distributing the CRL to vehicles. However, the disadvantage of relying only on passive revocation is the very nature of being passive. In other words, it leaves a time-gap between the time the system revokes a given vehicle and the time when the vehicle stops its communication. The latter occurs when either all certificates in the vehicle expires or it voluntarily stops communication as the result of rejection to request new certificates. 

For example, if a vehicle stores only one week worth of certificates, it likely requests the next set for the following one-week period sometime toward the end of the current period -- the timing in which the request is rejected if the vehicle is blocklisted. In worst-case scenario, this time gap can be up to 7 days. During this period, the vehicle continues to use its certificates; other vehicles certainly accept messages from this vehicle as valid. If the certificate reloading cycle becomes longer as the vehicle reload multiple weeks worth of certificates at a time, this time gap extends proportionally. Worse still, if the vehicle is preloaded with certificates for an extended period, such as 3 years, it does not request reloading for 3 years. This implies that certificate preloading for an extensive period is mutually exclusive with passive revocation approach. This way, the simplicity of the scheme comes at a price. ETSI ITS specification does not address this issue, thus requires further consideration.

\subsection{Certificate Management of RSUs}
\label{sec:issueUseofCertificates_ss5}
As discussed in Sec.~\ref{sec:issuePrivacyProtection_ss2}, RSUs do not require privacy protection and thus does not require the use of pseudonyms, strictly speaking. However, the certificate management system described in ETSI TS 102~940~\cite{etsi2021102940} does not address these types of ITS-S or if any specific certificate management different from privately owned vehicles is required at all. This is another open area that needs to be addressed in the standards. 

It may be necessary for vehicles to uniquely identify a specific RSU from another while still maintaining its anonymity. One possible approach is to assign pseudonyms with longer validity periods than those for vehicles, such as days, weeks, or months depending on how static the RSU pseudonyms can be. The similar principle may apply to non-privately-owned vehicles such as emergency vehicles. 
On the other hand, longer validity period negatively impacts revocation of such ITS-S types. For example, if an adversary hacks an RSU, steals its pseudonym certificates and corresponding private keys, and uses them on a fake RSU, this fake device can send legitimate messages for longer period than privately-owned vehicles. This situation is further pronounced in EU systems where it relies solely on \textit{passive revocation}. Therefore, a good balance needs to be achieved in order to minimize negative impacts from such situations. 

\subsection{Certificate Management of VRUs}
\label{sec:issueUseofCertificates_ss6}
The EU projects related to V2X communication define various use case scenarios \cite{naudts2018concordam19,fernandez20195gcard21,visintainer20195gcarmend21,perraud20195gcrocod21,martin20195gmobixd21}. These projects are under 5G-PPP, thus their technology focus is naturally on C-V2X as opposed to ITS-G5\@. Many of the use cases captured in these documents involve new and unique aspects in C-V2X compared to ITS-G5. One example is Vulnerable Road Users (VRU). VRU refers to pedestrians, cyclists, other human or non-human road users~\cite{etsi20191033001}. Inclusion of scenarios involving VRUs extends vehicular communication and contributes to further improvement of the road safety. 
In 2010, ETSI EN 302 665~\cite{etsi2010302665} defined handheld devices, or \textit{personal ITS-S}, as one of the ITS-S types. However, subsequent ETSI specifications focused exclusively on vehicle-centric view only. In this respect, increasing interest of VRUs due to the emergence of C-V2X was the trigger to start standardization work specific to VRUs. In fact, between 2019 and 2021, ETSI published TR 103 300-1~\cite{etsi20191033001}, TS 103 300-2~\cite{etsi20201033002}, and TS 103 300-3~\cite{etsi20211033003} which exclusively address VRU-related use cases, define functional architecture, and specify VRU basic service, respectively. The second specification~\cite{etsi20201033002} covers security-related issues. However, its content stays at the analysis level and leaves many issues open. The last one~\cite{etsi20211033003} specifies VRU Awareness Messages (VAM). It covers security aspects. However, the extent of its scope is limited; it does not define mechanisms and procedures such as enrolment of VRU devices and VRU-specific certificate policy including provisioning and usage of pseudonym certificates. In fact, it states that these areas are outside the scope of this specification (cf. clause 6.5.4 in~\cite{etsi20211033003}). 

The above situation also means that the VRUs (handheld devices) as a type of ITS-S were not originally envisioned as a part of the certificate management system defined in ETSI TS 102~940~\cite{etsi2021102940}, and it is still the case today. In fact, we have already pointed out in Sec.~\ref{sec:issuePrivacyProtection_ss2} that the existing management system~\cite{etsi2021102940} is strictly vehicle-centric view only. This situation raises several points: 1) inclusion of VRU as a type of ITS-S, including certificate management of VRUs, 2) definition of a certificate management back-end system for VRUs, equivalent for vehicles as in clause~7 in~\cite{etsi2021102940}. 

Including the VRU as a type of ITS-S has several implications. VRUs use smartphones to communicate with other ITS-S, such as indicating pedestrians' presence to nearby vehicles. This obviously means that smartphones, as a type of ITS-S, need to become legitimate members in the V2X communication system, including provisioning and usage of pseudonym certificates as in vehicle ITS-S. Because the coupling between a human user and his or her smartphone is even tighter than that of vehicles, even higher level of privacy protection is required for VRUs. In addition, if a specific VRU device needs to be revoked for any reason, appropriate mechanism needs to be in place to ensure that it is excluded from the V2X communication. Introduction of smartphones can serve as an easily-accessible potential new attack surface to the whole V2X communication system. As these consumer products are readily accessible than vehicle OBUs, the threshold is lower for adversaries to develop a malicious software on an open-source-based OS (e.g. Android) using available open-source software development tools -- picking up a smartphone and plugging in a USB cable to it is far easier and trivial than opening a part of a vehicle's dashboard, exposing a connection to the CAN bus and connecting to it. Hacking a vehicle OBU requires extensive knowledge, both mechanical and electronical, of vehicle's construction. In this sense, the above-mentioned VRU-specific specifications~\cite{etsi20191033001,etsi20201033002,etsi20211033003} 
fall short from addressing these aspects, leaving them as open issues.

One possible approach is to define a distinct ITS management system for smartphones separate from vehicles and RSUs. Separately manage smartphones likely simplifies the certificate management by isolating specific characteristics and aspects unique to them. One such example is the potential needs to interact with Mobile Network Operators (MNOs) if verification of the subscriber information is required before admitting the device as a legitimate member of the ITS system. At the same time, this separation of management also implies that an interconnection is needed between these two types of ITS-S management systems. In addition, multiple MNOs may be involved to accommodate subscribers of different MNOs, including MNOs of other countries to address roaming scenarios.

\subsection{Certificate Usage in Multiple Communications}
\label{sec:issueUseofCertificates_ss7}
It is likely that vehicles are engaged in multiple different types of communication with different entities for different purposes simultaneously. For example, a group of trucks in a platoon communicates with one another to coordinate their movement while maintaining safe driving distance with adjacent trucks within the platoon. In this case, these trucks likely use either unicast or multicast (groupcast) mode of communication rather than broadcast mode. At the same time, these trucks also broadcast basic service messages such as CAM to other surrounding non-platoon vehicles. The intended target and purposes of these messages are different.

In this case, it makes sense to use different pseudonyms for different purposes. This approach also aligns with the privacy protection perspective as one pseudonym used for broadcast mode does not reveal the pseudonym used for unicast mode. This way of pseudonym separation is likely to be beneficial, especially because broadcast mode does not provide confidentiality protection from observers (cf. Sec.~\ref{sec:issueCommModes_ss1}). Such separation of pseudonym usage enhances the security of V2X communication. 

ETSI specifications EN 302 636-1~\cite{etsi20143026361}, EN 302 636-3~\cite{etsi20143026363}, EN 302 636-4-1~\cite{etsi201730263641}, EN 302 636-6-1~\cite{etsi201430263661} discuss the use of multicast in GeoNetworking. They discuss security-related functionalities, such as authentication, authorization, integrity, privacy, and non-repudiation. However, they do not mention confidentiality especially in the context of user-plane traffic. Thus, encryption is not applied in multicast mode user traffic in GeoNetworking. 
Therefore, the vulnerabilities of multicast and broadcast traffic are at the same level. On the other hand, separate use of pseudonyms in unicast has a value as it can apply encryption, thus worth exploring this usage (cf.~Sec.~\ref{sec:issueCommModes_sss2_1}). However, it is not specified in these GeoNetworking related specifications and other key security-related specifications, such as TS 102~637-1~\cite{ts20101026371}, TS 102~940~\cite{etsi2021102940}, or TS 102~941~\cite{etsi2021102941}. As such, it embodies an interesting open challenge.

% !TEX root = v2x_survey_main.tex
% !TeX spellcheck = en-GB

% ========  Section 7  ==========

\section{Security Issue: Communication Modes}
\label{sec:issueCommModes}
\subsection{Broadcast-Oriented Communication}
\label{sec:issueCommModes_ss1}
The basic services provided by V2X communication includes CAM and DENM as defined in ETSI TS 302.637-2~\cite{etsi20193026372} and TS 302.637-3~\cite{etsi20193026373}, respectively. These messages are broadcast to the surrounding vehicles. Broadcast messages, by definition, are sent to any and all entities within the communication range. This is in contrast to unicast or multicast messages, which are sent to a specific endpoint or a known group of endpoints.
The very nature of the broadcast is that the transmitting node is not concerned with the number of receiving entities within the communication range and their identities. It is further aggravated with the dynamic topology changes due to moving vehicles. Therefore, a set of vehicles within a communication range of a vehicle are in the constant flux depending on the density and the relative speed differences. Another characteristic of broadcast messages is the absence of confidentiality protection as specified in TS 102.943~\cite{etsi2012ts102943}. Therefore, any entity with a suitable equipment can receive these messages.

These two points are significant from a V2X communication perspective because any passive observer with a suitable device can receive, collect, and analyse CAM and DENM messages. The receiving entity can verify the message authenticity and integrity by using the certificate contained in the message itself. Due to the use of periodically-changing pseudonyms and the unlinkability property from the privacy protection requirement, it is not trivial to identify the transmitting vehicle. However, a passive observer can still detect and recognize the existence of a specific pseudonym in the vicinity just by observing messages. This way, the broadcast nature of the basic messages has no or little protection from persistent observers to collect messages and detect long-term patterns of any given vehicle. This is an area of concern from a privacy protection perspective.

\subsection{Unicast – Confidentiality Protection}
\label{sec:issueCommModes_ss2}
\subsubsection{Applicability of Confidential Protection}
\label{sec:issueCommModes_sss2_1}
Confidentiality protection applies to unicast mode only. On the other hand, basic services are based on broadcast mode as discussed in Sec.~\ref{sec:issueCommModes_ss1}. This is captured in Table~2 in ETSI TR 102~893~\cite{etsi2017102893}. Unicast-based communication is rather a minority in V2X communication and is limited to specific use cases. In other words, confidentiality protection is applicable to a rather small portion of V2X communication where unicast is used. 
The 5GCAR D4.2 document in \cite{gallo20195gcard42} expresses a concern of the sole reliance on the PKI system \cite{adams2003PKI} for security and privacy of V2X communication. A proposed scheme in \cite{gallo20195gcard42} introduces a new entity called \textit{key manager} that generates symmetric keys for encryption. The proposed scheme is a step towards introducing an additional security mechanism. However, it overlooks the fact that the confidentiality protection is applicable to unicast mode only. Thus, it has limited applicability in V2X communication.  

\subsubsection{Usability of Unicast Communication}
\label{sec:issueCommModes_sss2_2}
The use of unicast in V2X communication is likely an IP-based communication to support value-added services. 
ETSI TS 102~941~\cite{etsi2021102941} states that the use of IPsec or TLS is assumed for the confidentiality protection in unicast. The use of IPsec or TLS implies a notion of a \textit{session} between two endpoints. Security Association (SA) establishment involves a handshake procedure which is an important factor to consider in a V2X communication environment where the vehicle topology changes constantly and dynamically.

The same characteristic of dynamic topology change as discussed in Sec.~\ref{sec:issueCommModes_ss1} equally applies to unicast communication. The communication range of DSRC is expected to be 300 meters~\cite{brecht2016security}. 
Then, depending on the relative speed and direction of vehicles, the communication between two ITS-Ss can be short-lived in the order of seconds. For example, if we assume communication between an RSU at a fixed location and a vehicle moving at 120 km per hour, the communication lasts only 9 seconds.
In the case of vehicles moving in the same direction, the communication between them may last longer. For example, if the relative speed difference of two vehicles is 10 km/hour, then the period they are within the communication range is 108 seconds, assuming a 300-meter range. Obviously, as the relative speed difference increases, the time duration shortens proportionally. Whether this time period is meaningful to establish an SA between two vehicles or not depends on the use cases and scenarios in which unicast communication is used. 

A prime example where the use of unicast makes sense is when one of the endpoints is a remote entity over the long-range communication (V2N), such as remote driving discussed in Sec.~\ref{sec:issuePrivacyProtection_ss5}. In this case, the vehicle’s location or speed is not a factor. However, this is a rather value-added use case beyond the basic service. Strictly in V2V scenarios, a group of trucks in a platoon is one example where unicast communication makes sense as they move in the same direction with short inter-vehicle distance for extended duration. However, in other situations of V2V or V2I communication, in the worst case, vehicles may go out of the communication range as soon as an SA is established, rendering SA establishment a moot point. It is worthwhile for the ETSI ITS specification to include a guidance on the usability of unicast in V2X communication.
% !TEX root = v2x_survey_main.tex
% !TeX spellcheck = en-GB

% ========  Section 8  ==========

\section{Security Issue: Message Handling}
\label{sec:issueMessageHandling}

\subsection{Plausibility Validation and Misbehavior Detection}
\label{sec:issueMessageHandling_ss3}

The concept of plausibility is present in different security aspects of V2X communications as specified in ETSI TR 102 893 \cite{etsi2017102893} and EN 302~636-4-1~\cite{etsi201730263641}. The idea of validation via plausibility, and why it is necessary in vehicular communication, is intuitively clear. A receiving vehicle needs to detect and reject bogus messages transmitted by an entity with a malicious intent to cause an accident or a road hazard, or to reject faulty messages transmitted by a vehicle with malfunctioning sensors. In this sense, plausibility validation is one of the key countermeasures to prevent potential threats in vehicular communication. There are two distinctive elements in the plausibility validation: 1) plausibility determination, and 2) misbehavior detection. 

\subsubsection{Determination of Plausibility}
\label{sec:issueMessageHandling_sss3_1}
ETSI TR 102~893~\cite{etsi2017102893} clause 11.3.20 states: \textit{“Plausibility checks are non-cryptographic measures which use rules and other mechanisms to determine the likelihood that received data is correct. These rules and mechanisms range from simple heuristics to quite sophisticated and more complex, methods.”} Also, clause B.4.5.3.2 in ETSI TR 102~893~\cite{etsi2017102893} describes suspicious behaviors as \textit{“any behavior that does not comply to expected behavior, based on direct evidence and probabilistic models.”} It lists examples such as spurious and bogus packets. A spurious packet contains a proper signature but flawed payload; a bogus message contains a flawed signature.
These definitions in the standard describe the intent of what plausibility check aims to accomplish, but it lacks clarity in terms of what qualifies as a good or valid plausibility check. Plausibility in V2X communication revolves around the idea of determining whether a given received message from another entity is reasonably genuine and thus should be accepted as valid. This level of message validation is above and beyond the verification of message integrity using digital signature.

Judging the plausibility of received messages may involve a number of contextual factors, such as: 1) location, 2) vehicle’s mobility (e.g.\ position, speed, and direction), 3) environment (e.g.\ local street or highway), 4) time of the day (e.g.\ rush hour or last night), and 5) other conditions (e.g.\ weather). 
However, these definitions in the standard are too abstract to be usable in reality to identify messages that do or do not conform to a given criterion. In other words, a more concrete and unambiguous definition is needed with respect to what constitute a set of criteria usable for plausibility validation. In addition, plausibility check needs to occur in real-time with high confidence to process time-critical messages such as DENM, which requires urgent and timely reaction such as an indication of an approaching emergency vehicle. In this sense, the result of plausibility validation determined too late or with less than 100 per cent confidence is either useless or may even result in an undesirable consequence.

Not defining a set of criteria for plausibility validation implies that it is left up to individual implementations on how it is accomplished. 
Such situation will most likely results in variations of coverage and effectiveness among them, where some of them better secure operational abilities than others in practice.  
Clearly, such situation is undesirable. However, establishing such criteria has a number of benefits such as:
\begin{enumerate}
\item Unambiguously defining how vehicles should behave under specific situations, 
\item Minimizing variation of implementations so that all vehicles on the road will have uniform and predictable behavior under the same situation,
\item Enabling OEMs to evaluate their implementations during development cycle for validation and improvement, 
\item Raising consumer confidence for successful adoption of the V2X technology in the market.
\end{enumerate}
While this is arguably a difficult area to standardize, possible definition of common criteria is worth pursuing in standards given the potential benefits. One possible approach is to define a common \textit{minimum set} of criteria that all OEMs must comply in their implementations. It can be done by defining a set of scenarios, associated with expected plausibility judgement and behaviour by vehicles. If a given OEM chooses to enhance its implementation with additional scenarios, it can be a differentiator from other OEMs without affecting the standardized set of scenarios.

\subsubsection{Misbehavior Detection}
\label{sec:issueMessageHandling_sss3_2}
Misbehavior detection is closely related to plausibility validation. They are two sides of the same coin. Detecting misbehavior implies detecting and analyzing patterns of plausibility validation failures on messages from another entity over time. Most likely, a single plausibility failure does not constitute a positive misbehavior detection. It requires continuous evaluation over time to determine if a misbehavior condition exists or not to reach reasonable level of confidence.
Studies by van der Heijden et al.\ \cite{vanderheijden2018survey} and Ambrosin et al.\ \cite{ambrosin2019design} provide good basis for various approaches toward misbehavior detection. Especially, the analysis in \cite{vanderheijden2018survey} captures a comprehensive overview and in-depth analysis of many misbehavior detection mechanisms. It analyzes and categorizes various mechanisms and classifies them into two dimensions, resulting in four categories.

The two dimensions described in \cite{vanderheijden2018survey} are: 1) \textit{node-centric vs. data-centric}, and 2) \textit{autonomous vs. collaborative}. 
The first dimension concerns with whether focusing on transmission patterns of a specific entity or analyzing data irrespective of the transmitting entity. The second dimension concerns with whether misbehavior detection is done locally within a node or as a result of exchanging information with other nodes.
The resulting four categories are:
\begin{itemize}
\item \textit{Behavioral} (\textit{node-centric} and \textit{autonomous})
\item \textit{Trust-based} (\textit{node-centric} and \textit{collaborative})
\item \textit{Plausibility} (\textit{data-centric} and \textit{autonomous})
\item \textit{Consistency} (\textit{data-centric} and \textit{collaborative})
\end{itemize}
Each of these four categories is based on certain conditions and assumptions. Thus, none of them is universally applicable under all circumstances. For example, \textit{collaborative}-based models (i.e.\ \textit{trust-based} and \textit{consistency}) imply that there are multiple vehicles in the area with which a vehicle can exchange information with to make an assessment on a specific vehicle. If there is not sufficient number of vehicles in the immediate area or not enough data is available to reach a conclusion, mechanisms in these categories are not effective. Another aspect of \textit{collaborative} category is the concept of \textit{honest majority} -- an assumption that the majority of the vehicles are honest and provide genuine information. However, if there is a small proportion of malicious entities or vehicles with faulty sensors providing incorrect data, it can skew the final decision on a vehicle in question.

The \textit{autonomous}-based models (i.e.\ \textit{behavioral} and \textit{plausibility}) solely rely on data within a vehicle. Therefore, the assessment and the decision of whether a given vehicle is misbehaving or not is necessarily limited to the information within the vehicle. The vehicle may reach a different decision if it has information other vehicles may have but does not locally. In this respect, combinations of all 4 categories is likely necessary to cover all possible scenarios and to gain confidence in the misbehavior detection. This is another area that needs further study. Ideally, a single solution that covers all possible situations is desirable. So far, the analysis in \cite{vanderheijden2018survey} indicates this is not the case.

Similar to plausibility validation, misbehaviour detection will likely result in different levels of effectiveness if it is left up to implementations. Defining a common approach or common test set to evaluate different implementations is worthwhile to guarantee uniform detection of misbehaviour.
As discussed in Sec.~\ref{sec:issueMessageHandling_sss3_1}, the same approach to standardize a common criteria will be helpful to achieve the same benefits for misbehaviour detection.

\subsection{Communication involving Vulnerable Road Users}
\label{sec:issueMessageHandling_sss3_3}

As discussed in Sec.~\ref{sec:issueUseofCertificates_ss6}, emergence of C-V2X introduces new use cases involving VRUs. They relate to the safety of road users other than vehicles such as pedestrians and cyclists who uses smartphones as a type of ITS-S. VRU-related messages include communicating presence and movement of pedestrians or cyclists to vehicles, or vice-versa. 
TS 102 300-3~\cite{etsi20211033003} specifies VAM; these messages enable VRU devices to communicate its position and movement with vehicles or other VRUs to improve road safety. Specifying these messages is a step forward to achieve this goal. However, this specification hints that more work is needed. Informative annex G in this specification gives a glimpse into unique issues and difficulties associated with VRU scenarios. It has to do with the unpredictable nature of pedestrian's movement compared to that of vehicles. One example is the transition from a pedestrian to a cyclist, and vice-versa, and how VAM messages accurately reflect this transition (e.g. VRU profile to change from a \textit{pedestrian} to a \textit{cyclist}). This annex indicates that these transitions are not a trivial problem to solve, thus requires further research and standardization.

% !TEX root = v2x_survey_main.tex
% !TeX spellcheck = en-GB

% ========  Section 9  ==========

\section{Security Issue: System Level Issues}
\label{sec:issueSystemLevel}

\subsection{ITS-S Device-Dependent Trust Level}
\label{sec:issueSystemLevel_ss2}
ITS-S consists of different types of devices. Vehicles are manufactured by OEMs and a large proportion of them is owned and used by private individuals. On the other hand, RSUs are special type of devices owned and managed by government authorities, and they are installed at fixed locations on a permanent basis. In this sense, it may make sense to differentiate the level of trust for vehicles and RSUs. In other words, the trust level for RSUs can be higher than vehicles and treat messages differently based on the message source. It implies that receiving vehicles need to distinguish message sources, at a coarse level such as \textit{RSU-type} or the \textit{vehicle-type}. Such differentiation does not compromise the privacy requirement (anonymity property). 

In addition, such device type-level identification has benefits. For example, trusting messages from RSUs may be useful in addressing new approaches in plausibility validation or misbehaviour detection discussed in Sec.~\ref{sec:issueMessageHandling_ss3}. One possible approach to address misbehaviour detection is to have intelligence in RSUs to collect and analyse data sent by vehicles, and integrate data from other RSUs at the back-end system to detect positive misbehaviour of vehicles. Then, this information can be distributed to vehicles in the area as an \textit{authoritative information}, which can only be sent by RSUs. A survey by Wang et al.\ \cite{wang2020certificate} discusses this concept as one of the approaches of certificate revocation by considering RSUs as an Intermediate Authority (IA). The concept of \textit{authoritative information} is somewhat akin to distributing CRL from the PKI system. 
The certificate definition in IEEE 1609.2~\cite{intelligent2016ieee16092} includes \texttt{SubjectAssurance} which can fit for this purpose. However, it also states that the exact content definition is outside its scope. This is another area for further research. 
	
A potential issue of this approach is an abuse of trust level if adversaries can compromise an RSU and modify its behaviour. However, protection against unauthorized tampering with malicious intent is a general issue that applies to all ITS-S types. Thus, it falls into the issue of hardening devices against such attacks. Specifically, TS 102 940~\cite{etsi2021102940} recommends the use of HSM as a solution.

\subsection{Interconnection between Multiple Security Management Systems}
\label{sec:issueSystemLevel_ss3}
A vehicular communication system involves government authorities that manage RSUs embedded in traffic lights and other road infrastructure that communicates with vehicles. This system also manages certificates to vehicles as specified in ETSI TS 102~940~\cite{etsi2021102940}. However, it is not just one management system, but multiple systems. A security management system most likely exists per appropriate geographical domain that has authority over specific jurisdiction. Depending on the deployment of the system, there may be one such management system at the national level, state or provincial level. It is up to individual country and its relevant authority to determine how this management system is owned and managed.

If we consider the certificate management system as described in ETSI TS 102~940~\cite{etsi2021102940}, each vehicle is expected to belong to one such management system. However, vehicles on the road likely belong to different such management systems. For example, highways in any given EU country are used by vehicles from multiple countries in addition to local vehicles. The above point implies that there needs to be an interaction between multiple management systems.

\begin{figure}[h]
  \centering
  \vspace{-0.2cm}
  \includegraphics[width=0.5\linewidth]{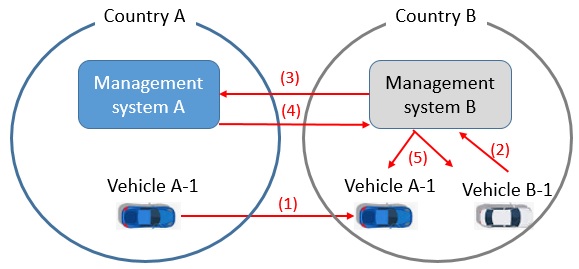}
  \caption{Scenario of Interconnection between Management Systems}
  \label{fig:interconnectionBetweenManagementSystems}
  \vspace{-0.3cm}
\end{figure}

The following examples illustrate the relevant scenarios. As the first example using Fig.~\ref{fig:interconnectionBetweenManagementSystems}, we consider a scenario where a vehicle ($V_{A1}$) from Country~A travels to Country~B (event~1). During its stay in Country~B, $V_{A1}$ is involved in a minor accident that causes some of the sensors to malfunction. As a result, $V_{A1}$ starts to report inaccurate or incorrect events and generates faulty messages to surrounding vehicles. In this case, a local vehicle ($V_{B1}$) in Country~B determines a misbehavior condition of $V_{A1}$ and reports this event to its certificate management system ($MS_B$) in Country~B (event~2). However, certificates of $V_{A1}$ were issued by the management system ($MS_A$) in Country~A. Therefore, there is nothing $MS_B$ can do unless there is an appropriate mechanism in place. It includes steps such as: 1) $MS_B$ to identify the $V_{A1}$’s country of origin and to notify such event to $MS_A$ (event 3), 2) $MS_A$ to revoke $V_{A1}$’s certificates and report it back to $MS_B$ (event 4).  

Another example is the active revocation of vehicles. This is applicable in the US system as discussed in Sec.~\ref{sec:systemOverview_sss3_2}. If the management system revokes a vehicle, it generates and distributes the CRL containing this vehicle’s information. However, in this case, vehicles ($V_{Bs}$) that encounter this vehicle on the road in Country~B also need to receive this CRL so that $V_{Bs}$ can correctly disregard any messages sent by $V_{A1}$. 
For this scenario to work correctly, $MS_B$ needs to be notified of the revocation condition of $V_{A1}$ from $MS_A$ so that $MS_B$ can distribute the CRL to vehicles in its territory. These aspects involving interaction across multiple security management systems are not covered in the ETSI ITS specifications. Given that cross-border mobility is a daily normal events in Europe, ETSI needs to address these aspects.

\subsection{Multiple Security Management Systems for Different ITS-S Types}
\label{sec:issueSystemLevel_ss4}
As discussed in Sec.~\ref{sec:issuePrivacyProtection_ss2}, ITS-S consists of multiple different types (i.e.\ vehicles, RSUs). However, ETSI TS 102~940~\cite{etsi2021102940} does not clearly specify how these different ITS-S types should be managed in the most effective and meaningful manner. Because these two types of ITS-S are owned and are used differently, it may make sense to manage them under separate management systems. For example, RSUs are owned, administered, and managed by government authorities, while majority of vehicles are owned and used by individual vehicle owners. One way to manage them is to manage RSUs under the national or provincial road authority, while vehicles are managed under regional vehicle registration offices. In addition, as discussed in Sec.~\ref{sec:issueUseofCertificates_ss6}, the inclusion of smartphones as a type of ITS-S in V2P scenarios raises a question of how they are managed to incorporate them as legitimate members of the ITS system. 
This is another aspect not currently covered by the ETSI ITS standard, thus it requires appropriate specification.

\subsection{Absence of Consideration on Post-Quantum Cryptography Technologies}
\label{sec:issueSystemLevel_ss5}
Both ETSI ITS~\cite{etsi2021103097} and IEEE 1609.2~\cite{intelligent2016ieee16092} use Elliptic Curve Cryptography (ECC) to generate digital signatures and encrypt messages. 
However, existing public key cryptographic algorithms, including ones based on ECC, are known to be vulnerable in the face of a quantum computer~\cite{shor1999polynomial}. 
Because vehicles, as an example of durable goods, have a lifespan as long as 23 years~\cite{oguchi2015regional}, they require technologies that can withstand against cyber threats during their lifetime. 
This includes migration to quantum-resistant security solutions~\cite{fernandez2019pre}. 
When vehicles adopt post-quantum (PQ) digital certificate in the future, a smooth transition from the conventional to new PQ certificates need to be ensured.

Consideration in this area involves two aspects in the V2X context: (1) support of post-quantum cryptography (PQC) technologies, and (2) new issues that stem from the support of such technologies. 
First, quantum computers are already a reality~\cite{ibmquantumcomputer}. Although it is expected to take many years for its capability to become an imminent threat~\cite{ericsson2021}, technologies being launched today need to have a solid migration strategy toward PQC paradigm. 
Evolving capabilities of quantum computers in the future necessitate cycles of updates in affected systems. 
This may imply that vehicles may require cycles of software update, upgrade, or even hardware replacement during the vehicle's lifetime to stay ahead of threats posed by future evolution of quantum computers. 
It further introduces a new challenge to securely execute these updates and replacements. 
This is an uncharted territory involving both vehicles and the underlying system infrastructure.
One such example is qSCMS~\cite{barreto2018qscms} which is a quantum-resistant version of butterfly key~\cite{simplicio2018unified}. 
Designing of new solutions to enhance or replace existing mechanisms to support PQ paradigm are required. 

Second, support of PQC means that the public key size will increase significantly from the conventional public key schemes. 
Since 2015, the US National Institute of Standard and Technology (NIST) started the process of selecting PQC~\cite{nistpostqccompetition}. 
If we assume a code-based quantum-resistant signature algorithm, the size of public key and signature from the ECC-based algorithm increase from 0.1KB to 190KB~\cite{kampanakis2018viability}. 
This will significantly impact the amount of storage space required in vehicles, especially if a large number of certificates are expected to be preloaded.  
This will make the preloading of 5 year worth of certificate as proposed in IFAL~\cite{verheul2019ifal} impractical, if not impossible.
This situation will further shift more toward the on-demand based AT reloading strategy we discussed in Sec.~\ref{sec:issueUseofCertificates_ss3} as a realistic solution.

% !TEX root = v2x_survey_main.tex
% !TeX spellcheck = en-GB

% ========  Section 10  ==========

\section{Root Cause Analysis And Recommendations}
\label{sec:rootCauseAnalysis}
Based on our analysis of ETSI ITS specifications, IEEE WAVE specifications, V2X related EU project documents, and relevant research papers, we have identified and discussed multiple gaps and issues in security aspects of V2X communication. 
We analyzed each of them and classified their origins into distinct root causes in Table~\ref{tab:summaryofSecurityIssues}. 
In this table, cells marked with a “\checkmark” indicate a gap in the applicable categories. 
All of these represent identified issues that need further study, research, and solutions.
In this section, we clarify each of their root cause category, along with our recommendations for the ETSI ITS specifications to address.
Table~\ref{tab:issueSummaryTbl} consolidates the identified issues.

\begin{table}[ht]
\footnotesize
\vspace{-0.3cm}
\caption{SUMMARY OF SECURITY ISSUES}
\vspace{-0.3cm}
\label{tab:summaryofSecurityIssues}
\begin{tabular}{llcccccccc}
	\hline
	\noalign{\hrule height 1pt}
	\rotatebox{90}{Root Cause Categories}		
	 	& \rotatebox{90}{Concerned Section} 
		& \rotatebox{90}{\begin{tabular}[c]{@{}l@{}}Conflicting or \\ insufficient specification\end{tabular}} 
		& \rotatebox{90}{\begin{tabular}[c]{@{}l@{}}Characteristics in vehicle \\ comm. environment\end{tabular}} 
		& \rotatebox{90}{\begin{tabular}[c]{@{}l@{}}Broadcast based \\ communication\end{tabular}} 
		& \rotatebox{90}{\begin{tabular}[c]{@{}l@{}}Implementation-dependent \\ ambiguity\end{tabular}} 
		& \rotatebox{90}{\begin{tabular}[c]{@{}l@{}}Certificate usage and \\ management for vehicles\end{tabular}} 
		& \rotatebox{90}{\begin{tabular}[c]{@{}l@{}}Certificate management \\ of non-vehicle ITS-S type\end{tabular}} 
		& \rotatebox{90}{\begin{tabular}[c]{@{}l@{}}RSU infrastructure \\ dependency\end{tabular}} 
		& \rotatebox{90}{\begin{tabular}[c]{@{}l@{}}MNO involvement \\ and dependency\end{tabular}} \\ \hline

	\noalign{\hrule height 1pt}
	\multicolumn{10}{c}{Sec.\ref{sec:issuePrivacyProtection} Privacy protection}                                                                                                                                                                                                                                                                                                                                                                                                                                                                                                                                                                                                                                                                                                                                                                                                                                                                                                                                                                               \\ \hline
	\begin{tabular}[c]{@{}l@{}}Privacy, Threat Actors, and Vehicle \\ Operation\end{tabular} 
		& \ref{sec:issuePrivacyProtection_ss0} & \checkmark & \checkmark  & \checkmark  & \checkmark  &  &   &   &  \\ \hline
	\begin{tabular}[c]{@{}l@{}}Applicability of Privacy 1: Vehicle \\Types and Usages\end{tabular} 
		& \ref{sec:issuePrivacyProtection_ss1} & \checkmark &   &   &   & \checkmark &   &   &  \\ \hline
	\begin{tabular}[c]{@{}l@{}}Applicability of Privacy 2: Non-Vehicle \\ITS-S\end{tabular} 
		& \ref{sec:issuePrivacyProtection_ss2} & \checkmark &   &   &   &   & \checkmark &   &  \\ \hline
	Privacy and Cooperative Awareness  
		& \ref{sec:issuePrivacyProtection_ss3} & \checkmark & \checkmark & \checkmark &   &   &   &   &  \\ \hline
	Privacy and Road Safety   
		& \ref{sec:issuePrivacyProtection_ss4} & \checkmark & \checkmark & \checkmark &   &   &   &   &  \\ \hline
	Privacy and Use of Unicast  
		& \ref{sec:issuePrivacyProtection_ss5} &   & \checkmark &   &   & \checkmark &   &   &  \\ \hline

	\noalign{\hrule height 1pt}
	\multicolumn{10}{c}{Sec.\ref{sec:issueUseofCertificates}: Use of certificate}                                                                                                                                                                                                                                                                                                                                                                                                                                                                                                                                                                                                                                                                                                                                                                                                                                                                                                                                                                            \\ \hline
	Certificate-based Message Verification  
		& \ref{sec:issueUseofCertificates_ss1} & \checkmark & \checkmark &   & \checkmark &   &   &   &   \\ \hline
	Certificate Usage and Change Policy  
		& \ref{sec:issueUseofCertificates_ss2} & \checkmark & \checkmark &   & \checkmark & \checkmark & \checkmark &   &   \\ \hline
	Certificate Reloading 
		& \ref{sec:issueUseofCertificates_ss3} & \checkmark &   &   & \checkmark & \checkmark &   & \checkmark &   \\ \hline
	Certificate Revocation 
		& \ref{sec:issueUseofCertificates_ss4} & \checkmark &   &   & \checkmark & \checkmark &   & \checkmark & \checkmark \\ \hline
	Certificate Management of RSUs 
		& \ref{sec:issueUseofCertificates_ss5} & \checkmark &   &   & \checkmark &   & \checkmark & \checkmark &   \\ \hline
	Certificate Management of VRUs
		& \ref{sec:issueUseofCertificates_ss6} & \checkmark &   &   & \checkmark &   & \checkmark &   & \checkmark \\ \hline
	Certificate Usage in Multiple Comm. 
		& \ref{sec:issueUseofCertificates_ss7} & \checkmark &   & \checkmark & \checkmark & \checkmark &   &   &   \\ \hline
	
	\noalign{\hrule height 1pt}
	\multicolumn{10}{c}{Sec.\ref{sec:issueCommModes}: Communication modes}                                                                                                                                                                                                                                                                                                                                                                                                                                                                                                                                                                                                                                                                                                                                                                                                                                                                                                                                                                           \\ \hline
	Broadcast-Oriented Communication 
		& \ref{sec:issueCommModes_ss1}   &   & \checkmark & \checkmark &   &   &   &   &   \\ \hline
	Unicast – Confidentiality Protection 
		& \ref{sec:issueCommModes_ss2}   &   & \checkmark &   &   &   &   &   &   \\ \hline

	\noalign{\hrule height 1pt}
	\multicolumn{10}{c}{Sec.\ref{sec:issueMessageHandling}: Message handling}                                                                                                                                                                                                                                                                                                                                                                                                                                                                                                                                                                                                                                                                                                                                                                                                                                                                                                                                                                             \\ \hline
	\begin{tabular}[c]{@{}l@{}}Plausibility Validation and \\Misbehavior Detection\end{tabular}  
		& \ref{sec:issueMessageHandling_ss3}     & \checkmark & \checkmark &   & \checkmark &   &   &   &    \\ \hline
	Communication involving VRU  
		& \ref{sec:issueMessageHandling_sss3_3}  & \checkmark &   &   &   &   & \checkmark &   & \checkmark  \\ \hline

	\noalign{\hrule height 1pt}
	\multicolumn{10}{c}{Sec.\ref{sec:issueSystemLevel}: System level}                                                                                                                                                                                                                                                                                                                                                                                                                                                                                                                                                                                                                                                                                                                                                                                                                                                                                                                                                                                   \\ \hline
	ITS-S Device-dependent Trust Level   
		& \ref{sec:issueSystemLevel_ss2}    & \checkmark  &   & \checkmark &   &   &   & \checkmark &   \\ \hline
	\begin{tabular}[c]{@{}l@{}}Interconnection between Multiple \\ Security Management Systems\end{tabular}  
		& \ref{sec:issueSystemLevel_ss3}    & \checkmark  &   &   & \checkmark & \checkmark &   &   &   \\ \hline
	\begin{tabular}[c]{@{}l@{}}Multiple Security Management \\Systems for Different ITS-S Types\end{tabular} 
		& \ref{sec:issueSystemLevel_ss4}    & \checkmark  &   &   & \checkmark & \checkmark & \checkmark &   & \checkmark \\ \hline
	\begin{tabular}[c]{@{}l@{}}Absence of Post-Quantum \\Cryptography Technologies \end{tabular}   
		& \ref{sec:issueSystemLevel_ss5}    & \checkmark  &   &   &   &   &   &   &   \\ \hline
	\noalign{\hrule height 1pt}
\end{tabular}
\end{table}

\par\noindent\textbf{Root cause 1: Conflicting or Insufficient Specification.} We have identified a number of areas where existing ETSI ITS specifications contain conflicting requirements and areas that are not sufficiently specified. 
Conflicting requirements likely lead to an ineffective system as the end result. 
Insufficient specification means certain security aspects are specified for a certain subset only, or implicitly left to implementation decisions.

\noindent\textit{Recommendations \& research objectives: } 
\begin{itemize}
  \item{
Solve the conflicting requirements between privacy, anonymity, and safety
-- this refers to the situation where the mechanisms to ensure privacy of a vehicle owner result in potential compromise in road safety.
One example is the unobservability property making vehicles not being able to determine whether to apply brake or not upon receiving EEBL message from a surrounding vehicle (Sec.~\ref{sec:issuePrivacyProtection_ss4}). Another example is the certificate management of non-vehicle ITS-S types (Sec.~\ref{sec:issuePrivacyProtection_ss2}, ~\ref{sec:issueUseofCertificates_ss5},~\ref{sec:issueUseofCertificates_ss6}, and ~\ref{sec:issueMessageHandling_sss3_3}).
}
  \item{
Clarify the relationship and define the interworking mechanism between
security management systems of different organizational entities,
jurisdictions, and countries -- vehicles from multiple regions and countries need to be able to communicate seamlessly across boundaries. This includes validation of pseudonym certificates in the receives messages, plausibility validation, and misbehavior detection (Sec.~\ref{sec:issueMessageHandling_ss3} and~\ref{sec:issueSystemLevel_ss3})\@. 
}
  \item{The integration and joint risk
assessment of complex system requirements that encompass safety, security,
and privacy is a poorly understood field of research. Analysis techniques
that combine approaches from safety engineering with those from security
engineering, e.g. STPA-Sec or integrating attack vectors with fault trees
may help but have not been applied to systems at the scale of V2X. }
\end{itemize}

\par\noindent\textbf{Root cause 2: Characteristics of Vehicular Communication Environment.} The fundamental characteristic of a vehicle in operation is its dynamically changing position, both in relation to a static geo-referential or other vehicles in operation. The paradigm of constant topological changes combined with short-range direct communication poses challenges. 
This also applies to GeoNetworking as it is a chain of short-range communication between vehicles. 
This fundamental characteristic likely limits the type of services that can be realized in such environment. In particular, direct communications that are expected to last longer or between specific endpoints in unicast mode may suffer from a communication loss and impact its services as a result.

\noindent\textit{Recommendations \& research objectives:} 

\begin{itemize}
  \item{
Establish realistic expectations of the communication in both broadcast and
unicast modes. Articulate the criteria and condition to use unicast communication as vehicles at the borderline of
communication range will likely suffer communication failure
(Sec.~\ref{sec:issuePrivacyProtection_ss5}, ~\ref{sec:issueCommModes_ss2}).
}
  \item{Establish guidelines on how to use and change pseudonyms for different purposes effectively in dynamically changing topology (Sec.~\ref{sec:issueUseofCertificates_ss2}, ~\ref{sec:issueUseofCertificates_ss7}). 
}
  \item{A potential avenue towards increasing
fault tolerance of V2X networks is to improve peer-to-peer
networking between road users. To improve road safety, V2X
technology must be used together with direct sensor perceptions in autonomous vehicles.}
\end{itemize}

\par\noindent\textbf{Root cause 3: Broadcast Based Communication.}
Basic services in the V2X are CAM and DENM which
are broadcast based communication. Its fundamental characteristics are
that: 1) any entity can receive messages, and 2) confidentiality protection
is not applied. These two points render CAM and DENM \textit{open}
communication. There are solutions to apply encryption to broadcast traffic~\cite{du2005id,garay2000long,sakai2007identity,halevy2002lsd}.
However, the ETSI ITS standard does not require such mechanism. This
situation makes it trivial for a malicious entity to eavesdrop and
collect data (cf. Sec.~\ref{sec:issueCommModes_ss1}). The only solution to render vehicular
communication trustworthy is to apply the PKI-based message authentication and validation mechanism~\cite{adams2003PKI}. 
This makes V2X infrastructure dependent on security and reliability of PKI systems. 

\noindent\textit{Recommendations \& research objectives:} 
\begin{itemize}
  \item{
Consider alternative approaches to apply confidentiality protection or
pseudonymity to the
broadcast-based communication, investigate efficient means for
cryptographic credential and trust management at scale. 
}
\end{itemize}

\par\noindent\textbf{Root cause 4: Implementation-Dependent Ambiguity.} We
discussed the challenges, such as CRL distribution, plausibility validation, and misbehaviour detection. These areas are
often implementation specific and are left up to the individual
OEM’s decision. This situation results in variations of effectiveness
among implementations, where one implementation performs better than others in real life. 
It is likely not realistic to specify every aspects of the communication system in the standard. In fact, it may make sense or is inevitable to leave some aspects to implementation choice. However, those aspects should at least be documented in the standard to be cognizant of the choice of the extent the standard does not cover. This way, possible consequences are made clear. 

\noindent\textit{Recommendations \& research objectives:}
\begin{itemize}
  \item{
Define baseline criteria, which consists of a set of test cases, test data, and resulting decision criteria as a common foundation upon which various implementations can be tested and evaluated against. Doing so has a number of benefits (Sec.~\ref{sec:issueMessageHandling_sss3_1}). One example is to define a clear and unambiguous method to validate plausibility and detect misbehaviour (Sec.~\ref{sec:issueMessageHandling_ss3}).
}
\end{itemize}

\par\noindent\textbf{Root cause 5: Certificate Usage and Management for Vehicles.} Vehicles' usage of pseudonym certificates is not well defined in ETSI ITS specifications. This includes their usage period, change rules, and reloading mechanisms. In addition, another unspecified area is the pseudonym usage for different purposes (e.g. broadcast and unicast).

\noindent\textit{Recommendations \& research objectives:}
\begin{itemize}
  \item{
Define how pseudonym and certificate are used by vehicles. This includes
usage and change rules (Sec.~\ref{sec:issueUseofCertificates_ss2}),
revocation and reloading rules and policy (Sec.~\ref{sec:issueUseofCertificates_ss3},~\ref{sec:issueUseofCertificates_ss7}).
}
  \item{Investigate efficient means for
cryptographic credential and trust management at scale, consider the use of
light-weight Hardware Security Modules (HSM) and Trusted Execution Environments (TEE)
in vehicles.}
\end{itemize}

\par\noindent\textbf{Root cause 6: Certificate Management of Non-Vehicle ITS-S Types.} Pseudonym usage by RSUs is not explicitly specified in the ETSI ITS standard. Thus it remains unclear if and how their usage is different from vehicles. Introducing smartphones (VRU) as a type of ITS-S requires appropriate management of these devices. Smartphones are consumer-owned generic platform as opposed to dedicated purpose devices such as OBUs and RSUs, thus a different approach to manage them will be required. 
Another open issue of smartphones is whether and how they can be revoked and removed from ITS system, if and when it is necessary.

\noindent\textit{Recommendations \& research objectives:}
\begin{itemize}
  \item{
Define how the pseudonym usage in RSU is different from vehicles (Sec.~\ref{sec:issueUseofCertificates_ss5}). Define how smartphones as a type of ITS-S is managed in the ITS system, including enrolment, verification, and authorization, issuance of certificates, usage of pseudonyms, and how they are revoked from the system (Sec.~\ref{sec:issueUseofCertificates_ss6}).
}
\end{itemize}

\par\noindent\textbf{Root cause 7: RSU Infrastructure Dependency.} Many functionalities discussed in EU project use cases depend on RSUs and the infrastructure behind them, such as distribution and reloading of certificates. However, ubiquitous installation and availability of RSUs is an assumption, not a given condition. How soon an RSU infrastructure will be deployed depends on a number of non-technical factors, such as government policy on ITS and budget allocation. It is likely a gradual process and varies from one region to another and one country to another. ITS-capable vehicles likely find themselves in the situation where RSU installation is either scarce or non-existent. Thus, being overly dependent on RSUs is counterproductive to the deployment of V2X technology.

\noindent\textit{Recommendations \& research objectives:}
\begin{itemize}
  \item{
Define alternative solutions to reduce dependency on the ubiquitous RSU deployment in such a way that necessary functionalities can be fulfilled independent from RSUs when it is necessary, while making effective use of RSUs when they are available (Sec.~\ref{sec:issueUseofCertificates_ss3}).
}
\end{itemize}

\par\noindent\textbf{Root cause 8: MNO Involvement and Dependency.} Within
the context of VRUs, whether interaction is required  between the ITS
system and MNOs to manage their access to the ITS system is unspecified in
the ETSI ITS specification. If required, the solution needs standardization so that it will be adopted by all MNOs at international level. 

\noindent\textit{Recommendations \& research objectives:}
\begin{itemize}
  \item{
Define if the ITS management system and MNOs need to interact with each other to manage VRUs (smartphones) or not. This  interaction between them includes definition of message contents (Sec.~\ref{sec:issueUseofCertificates_ss6}). This involves coordination with appropriate standard bodies such as 3GPP.
}
\end{itemize}

\begin{table}[h]
	\footnotesize
	\vspace{-0.2cm}
	\caption{LIST OF RECOMMENDED ACTIONS TO ETSI ITS SPECIFICATIONS}
	\vspace{-0.3cm}
	\label{tab:issueSummaryTbl}
	\begin{tabular}{llll}
		\hline
			\noalign{\hrule height 1pt}
		Num & Description & Section \\ \hline
		\noalign{\hrule height 1pt}
		1 & \begin{tabular}[c]{@{}l@{}}Specify possible differences in the privacy protection requirements for different ITS-S types.  \end{tabular} & \ref{sec:issuePrivacyProtection_ss1},~\ref{sec:issuePrivacyProtection_ss2} \\ \hline
		2 & \begin{tabular}[c]{@{}l@{}}Resolve conflicting requirements of privacy protection and including vehcile-identifiable \\information in CAM and DENM.  \end{tabular} & \ref{sec:issuePrivacyProtection_ss3} \\ \hline
		3 & \begin{tabular}[c]{@{}l@{}}Resolve conflicting requirements of privacy protection and road safety. \end{tabular} & \ref{sec:issuePrivacyProtection_ss0}, \ref{sec:issuePrivacyProtection_ss4} \\ \hline
		4 & \begin{tabular}[c]{@{}l@{}}Resolve privacy protection in unicast communication from both internal and external threats. \end{tabular} & \ref{sec:issuePrivacyProtection_ss5} \\ \hline
		5 & \begin{tabular}[c]{@{}l@{}}Address the potential issue of real-time certificate validation. \\  \end{tabular} & \ref{sec:issueUseofCertificates_ss1} \\ \hline
		6 & \begin{tabular}[c]{@{}l@{}}Standardise certificate change and reloading rule or policy. \\  \end{tabular} & \ref{sec:issueUseofCertificates_ss2}, \ref{sec:issueUseofCertificates_ss3}\\ \hline
		7 & \begin{tabular}[c]{@{}l@{}}Resolve the time gap associated with passive revocation. \\  \end{tabular} & \ref{sec:issueUseofCertificates_ss4} \\ \hline
		8 & \begin{tabular}[c]{@{}l@{}}Specify certificate management of RSUs and VRUs. \\  \end{tabular} & \ref{sec:issueUseofCertificates_ss5}, \ref{sec:issueUseofCertificates_ss6} \\ \hline
		9 & \begin{tabular}[c]{@{}l@{}}Specify different use of certificates for different purposes. \\  \end{tabular} & \ref{sec:issueUseofCertificates_ss7} \\ \hline
		10 & \begin{tabular}[c]{@{}l@{}}Address vulnerabilities of broadcast messages against passive observers. \\  \end{tabular} & \ref{sec:issueCommModes_ss1} \\ \hline
		11 & \begin{tabular}[c]{@{}l@{}}Establish guidances on the use of unicast mode. \\  \end{tabular} & \ref{sec:issueCommModes_ss2} \\ \hline
		12 & \begin{tabular}[c]{@{}l@{}}Establish guidance on the implementation of plausibility validation and misbehaviour detection.  \end{tabular} & \ref{sec:issueMessageHandling_ss3} \\ \hline
		13 & \begin{tabular}[c]{@{}l@{}} Investigate solutions for accurate detection of VRU movement to VAM.  \end{tabular} & \ref{sec:issueMessageHandling_sss3_3} \\ \hline
		14 & \begin{tabular}[c]{@{}l@{}}Consider possible notion of device type-dependent trust level. \\  \end{tabular} & \ref{sec:issueSystemLevel_ss2} \\ \hline
		15 & \begin{tabular}[c]{@{}l@{}}Analyse and establish scenarios of interconnecting multiple management systems. \\  \end{tabular} & \ref{sec:issueSystemLevel_ss3},~\ref{sec:issueSystemLevel_ss4} \\ \hline
	\end{tabular}
\end{table}

\section{Conclusion and Future Work}
\label{sec:conclusion}
In this paper, we examined the EU standards for ITS, related US standards, various V2X-related EU projects, and relevant research papers. We integrated information from these sources, analyzed, and identified gaps in the security aspects of vehicular communication, focusing on the ETSI ITS specifications.

The issues and gaps we have identified are significant. They include conflicting and undefined specification in the standards. System level aspects need further definition, e.g. interworking of multiple management systems across multiple jurisdictions and countries. The security management of VRUs (smartphones) as a new ITS-S type is missing. Without addressing these aspects, the vehicular communication in reality can very well be unreliable, insecure, and unusable, ultimately leading to accidents, injuries, or loss of properties. Leaving details to implementation-specific solutions can lead to varying degree of effectiveness among implementations. To ensure uniform operation and effectiveness, further research and standardization is needed. In this respect, the future work is to address these gaps and issues to define possible solutions.

As a conclusion, the security solution in the ITS standards solely based on PKI leaves a number of areas to reconsider. Additional approaches and solutions are required to ensure vehicular communication is indeed secure so that the overall objective to make roads safer and reduce road accidents can be achieved rather than providing a new target for cyberattacks. The fundamental nature of cyber-physical system, such as vehicular communication, is that a sub-optimal system can cause physical damage in reality. In order to avoid such losses, all relevant security and privacy aspects of vehicular communication need to be addressed.

%% 
%\section{Citations and Bibliographies}
\label{sec:citations}

%%
%% The acknowledgments section is defined using the "acks" environment
%% (and NOT an unnumbered section). This ensures the proper
%% identification of the section in the article metadata, and the
%% consistent spelling of the heading.
%\vspace{-5pt}
\begin{acks}
This work was supported in part by CyberSecurity Research Flanders with
reference number VR20192203. This work was also supported in part by the
Research Council KU Leuven C1 on Security and Privacy for Cyber-Physical
Systems and the Internet of Things with contract number C16/15/058. In
addition, this work was supported in part by the Flemish Government through
the imec Netsec project, through the EIT Health RAMSES project, through the
Smart Highways SErVo project, and by the European Commission through the
Horizon 2020 research and innovation programme under grant agreement
H2020-SC1-FA-DTS-2018-1-826284 ProTego and MSCA-ITN-814035 5GhOSTS.
\end{acks}
%\vspace{-5pt}
%%
%% The next two lines define the bibliography style to be used, and
%% the bibliography file.
\bibliographystyle{ACM-Reference-Format}
%\bibliography{sample-base}
\bibliography{bibliography_1}

%%
%% If your work has an appendix, this is the place to put it.
\appendix

\end{document}